\documentclass[fleqn,usenatbib,apj,twocolappendix,twocolumn]{openjournal}
\shorttitle{Multiphase CGM turbulence}
\shortauthors{Mohapatra et al.}

\usepackage{graphicx}	
\usepackage{amsmath}	

\usepackage[breaklinks,color links,citecolor=blue,urlcolor=blue]{hyperref}
\hypersetup{colorlinks=true,linkcolor=blue,citecolor=blue,filecolor=blue,urlcolor=blue}
\usepackage{natbib}
\usepackage{ragged2e}
\usepackage{comment}
\usepackage{capt-of}
\usepackage{bigints} 
\usepackage{physics}
\usepackage{xcolor}
\usepackage{cleveref}
\usepackage{longtable}
\usepackage{bm}
\usepackage{cancel}
\usepackage{comment}
\usepackage{rotating}

\allowdisplaybreaks

\DeclareRobustCommand{\orcidauthor}[3]{%
  \author{%
    \href{https://orcid.org/#1}{#2%
      \IfFileExists{orcid-ID.png}{\ \raisebox{-0.2ex}{\includegraphics[width=9pt]{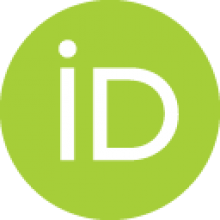}}}{}%
    }\textsuperscript{#3}%
  }%
}

\def\mean#1{\left< #1 \right>}
\def\sun{\odot}

\begin{document}
\title[Multiphase CGM turbulence]{Multiphase gas in Circumgalactic cloud complexes: Insights from kiloparsec-scale Magnetohydrodynamic Turbulence Simulations}


\orcidauthor{0000-0002-1600-7552}{Rajsekhar Mohapatra}{1,*}
\affiliation{$^{1}$Department of Astrophysical Sciences, Princeton University, Princeton, NJ 08544, USA}

\orcidauthor{0000-0002-9287-4033}{Alankar Dutta}{2}
\affiliation{$^{2}$Max-Planck-Institut f\"ur Astrophysik, Karl-Schwarzschild-Str. 1, 85748 Garching b. M\"unchen, Germany}
\orcidauthor{0000-0003-2635-4643}{Prateek Sharma}{3}
\affiliation{$^{3}$Department of Physics, Indian Institute of Science, Bangalore, KA 560012, India}

\email{* rmohapatra@princeton.edu} 

\begin{abstract}
The circumgalactic medium (CGM) is the diffuse gas surrounding a galaxy's halo, and it plays a vital role in the galactic baryon cycle. However, its mass distribution across the virial hot phase, and cooler and denser atomic and molecular phases remains uncertain, complicating our understanding of galaxy evolution.
To investigate this, we perform high-resolution magnetohydrodynamic simulations of 0.125--8 kpc-scale representative patches of the CGM, with parameters informed by quasar absorption line observations. Our simulations employ global thermal balance and resolve the cooling length (the minimum across all temperatures of $c_s t_{\rm cool}$, where $c_s$ is the sound speed and $t_{\rm cool}$ is the cooling time in isobaric conditions), allowing us to track the evolution of cold gas more accurately. We find that low-density CGM gas ($3\times10^{-4}$ cm$^{-3}$) cannot sustain cold gas below $10^4$ K for long, due to a large value of the ratio between the cooling to mixing time ($t_{\rm cool}/t_{\rm mix}$). In contrast, higher-density environments ($3\times10^{-3}~{\rm cm}^{-3}$) reach a turbulent multiphase steady state, with up to $50\%$ of the mass in the cold phase, occupying only about $1\%$ of the volume. To connect with large-volume cosmological simulations and small ${\rm pc}$-scale idealized simulations, we explore different box sizes (0.125--8 kpc) and identify a key scaling relation: simulations with similar $t_{\rm cool}/t_{\rm mix}$ exhibit comparable cold gas mass fractions and lifetimes. Importantly, we find that simply sub-sampling (reducing box-size) a small region from a large-volume simulation while maintaining a constant turbulent energy density injection rate from larger to smaller scales artificially shortens $t_\mathrm{mix}$, leading to inaccurate predictions for cold gas survival. This means that cold gas at small $\lesssim 10$ kpc scales arises in relatively dense, quiescent regions of the CGM rather than the turbulent ones undergoing cascade from large scales. 
\end{abstract}



\section{Introduction}\label{sec:introduction}
The circumgalactic medium (CGM) refers to the gas pervading the halo of a galaxy. It is expected to contain similar, if not larger baryonic mass compared to the galaxy \citep{Tumlinson2017review} and plays a critical role in the galactic baryon cycle. The CGM interacts with the galaxy through gas inflow and outflow. For example, it interacts with the galactic wind and fountain flows, and mediates the galaxy's interaction with the intergalactic medium and other galaxies. Although most of the volume-filling component of the CGM is expected to be hot and ionized, at temperatures close to the virial temperature of the halo, substantial amounts of colder and denser atomic and molecular gas have been inferred \citep{Lehner2015ApJ,Borthakur2015ApJ}. 

Due to its low density, both the hot ($T\gtrsim10^6~\mathrm{K}$ for a Milky Way mass galaxy) and cold components (at $T\lesssim10^4~\mathrm{K}$) of the CGM have usually been difficult to detect in emission. The hot virial component CGM has been characterized by stacked X-ray and thermal Sunyaev-Zeldovich effect \citep{Singh2018MNRAS,Zhang2024A&A,Das2023ApJ} measurements. The cooler CGM, on the other hand, has been studied in absorption through detailed photoionization modeling of the absorption profiles of foreground CGM on distant quasar spectra \citep{Tumlinson2013ApJ,Werk2014ApJ,HWChen2020MNRAS}. For the Milky Way, there is also some evidence for multiple sub-components of the hot phase \citep{Das2021ApJ,Lara2023ApJ} at sub and super-virial temperatures. 
Other methods, such as the dispersion and scattering of distant FRBs, have also been used/proposed to characterize the properties of the CGM and its microphysical structure \cite{Ravi2019ApJ,Ocker2025ApJ}. 

On the analytic modeling front, studies such as \cite{Faerman2017ApJ,Faerman2020ApJ} have proposed models of the CGM constrained by observations of its warm and hot components. \cite{Voit2019ApJ} took into account the precipitation limit of the hot CGM to obtain further constraints on and obtain predictions for the highly ionized CGM absorption lines such as \texttt{OVI, OVII, OVIII, NeVIII, NV}. \cite{Singh2024MNRAS} present a comparison between the different CGM models such as isoentropic, precipitation-limited, cooling-flow, etc. While most models so far have focused on the volume-filling hot phase, a few recent models have also incorporated cold clouds using analytic/geometric models; examples include CloudFlex \citep{Hummels2024ApJ}, modeling the CGM using log-normal distributions \cite{ADutta2024MNRASlognormal}, and a combination of dense cloud-complexes and hot CGM \citep{Bisht2025MNRAS}.

On the numerical modeling front, the CGM and its components have been studied using cosmological simulations and zoom-ins \citep{Nelson2020MNRAS,Hafen2020MNRAS}, with some recent studies using special refinement schemes to better resolve the cooler components \citep{Ramesh2024MNRAS,Rey2024MNRAS}. At smaller scales, several studies have looked into the interaction of individual clouds interacting with a hot background \citep[][to name a few]{Scannapieco2015ApJ,Armillotta2016,Gronke2018,Kanjilal2021MNRAS,Jung2023MNRAS,Dutta2025MNRAS} and filaments \citep{Berlok2019MNRAS,Mandelker2020MNRAS,Kaul2025MNRAS}. These studies have obtained constraints on the survival of cold clouds as they interact with the hot background, obtaining size limits on surviving clouds and how it can be affected by magnetic fields, density contrasts with the hot phase, etc. 
On even smaller scales, \cite{Fielding2020ApJ,Tan2021MNRAS,Zhao2023MNRAS,Das2024MNRAS} have zoomed in to the turbulent radiative mixing layer (TRML) at the interface between the cold and hot phases, ubiquitous for most of these multiphase media. They have demonstrated that the TRML has a fractal structure, and obtained estimates of mass and energy flux between the hot and cold phases. Resolving the detailed temperature structure of the mixing layer is also useful for obtaining constraints on the density and temperature of the cold gas from the observed fluxes of intermediate temperature ions such as \texttt{CIV, SiIV OV, OVI}, etc.~which have a short recombination time. 

At scales intermediate between the large-volume cosmological simulations and the small-volume cloud/filament-wind interactions and TRML simulations, there are another set of meso-scale studies that model a small patch of the CGM as a turbulent box. These studies, for e.g. \cite{Buie2018ApJ,Mohapatra2022MNRASb,Gronke2022MNRAS,Fielding2023ApJ,Das2024MNRAS}, are agnostic to the driver of turbulence—for the CGM it could be Active Galactic Nuclei (AGNs) jets, galactic outflows, or sloshing due to mergers, etc. Instead, they assume that as long as turbulence is fully developed and we study the CGM at scales much smaller than the turbulence driving scale, the statistical properties of the CGM at the box size scale can be well-represented by a turbulent box with similar physical properties. Although these simulations are meso-volume, with box sizes ranging from few $(\mathrm{kpc})^3$ to few $(100\mathrm{kpc})^3$, they can be useful to study both physical and statistical properties of the CGM. Some example studies include the survival of cold clouds in a multiphase turbulent medium \citep{Gronke2022MNRAS,Ghosh2025arXiv}, the relations between hot and cold phase gas velocities \citep{Mohapatra2022MNRASa}, the mass and size-distribution of cold clouds \citep{Fielding2023ApJ}.

Turbulence in the CGM can be driven by interactions with satellite substructures, as well as by galactic outflows powered by supernovae and AGN feedback. It plays a pivotal role in regulating the thermodynamic evolution of the CGM. In particular, turbulence influences the precipitation limit \citep{Voit2018ApJ,Mohapatra2023MNRAS,Wibking2025MNRAS}, which governs the onset of multiphase condensation: when the gas cooling time falls below a threshold set by this limit, cold gas can condense out of the hot phase. Additionally, turbulence affects the longevity of cold gas embedded within the hot medium, as demonstrated by \citet{Gronke2022MNRAS}. Observational signatures of CGM turbulence have been identified through absorption-line studies at low redshift \citep{HWChen2023ApJ}, as well as through emission from extended Quasi Stellar Object (QSO) nebulae \citep{MChen2024ApJ,MChen2025ApJ}.

Despite its importance and observational evidence, there remains a lack of targeted simulations exploring the parameter space of CGM turbulence while simultaneously resolving key physical scales, such as the minimum cooling length of cold gas, and small scales where turbulent diffusion dominates over numerical mixing of cold and hot gas. In this work, we address this gap using high-resolution magnetohydrodynamic simulations of dense multiphase cloud complexes in the CGM, with physical parameters directly informed by QSO absorption line measurements. We perform meso-scale simulations of turbulence in these cloud complexes with box sizes ranging from $0.125$—$8~\mathrm{kpc}$. Our simulations resolve the cooling length of the gas at all densities, by at least a factor of $10$ for gas at temperatures larger than $10^4~\mathrm{K}$. We have also verified that multiphase mixing in our simulations is not strongly affected by numerical diffusion. Further, for turbulence cascade in a uniform box, the mixing time scales as $l^{2/3}$ \citep[assuming Kolmogorov scaling][]{kolmogorov1941dissipation}, implying that the smaller-scale multiphase gas is mixed rather than being sustained by cooling, since the cooling time is scale-independent. We investigate the importance of this effect with two sets of simulations, one where we maintain a constant energy cascade rate and the other with a constant mixing time across scales. The latter regime is needed to sustain coherent cold gas at inferred at $\lesssim 10$ kpc in observations (\citealt{Rudie2019,Augustin2023,Afruni2023}).

This paper is organized as follows. In \S\ref{sec:Methods}, we describe our simulation setup and numerical methods. Key results from our simulations are presented in \S\ref{sec:results}, followed by a discussion of their implications in a global CGM context in \S\ref{sec:varying_box_size}. In \S\ref{sec:sensitiveness_to_sim_params}, we examine how variations in simulation parameters affect our findings. Limitations of the current study and directions for future work are discussed in \S\ref{sec:Caveats_future_work}. Finally, we summarize and conclude in \S\ref{sec:summary_discussion}.

\section{Methods}\label{sec:Methods}

\subsection{Simulated Equations}\label{subsec:ModEq}

We model the CGM as a fluid governed by the compressible magnetohydrodynamic (MHD) equations, along with the ideal gas equation of state. The system evolves according to:

\begin{subequations}
	\begin{align}
	\label{eq:continuity}
	&\frac{\partial\rho}{\partial t}+\nabla\cdot (\rho \mathbf{v})=0,\\
	\label{eq:momentum}
	&\frac{\partial(\rho\mathbf{v})}{\partial t}+\nabla\cdot (\rho \mathbf{v}\otimes \mathbf{v}+P^*I -\mathbf{B}\otimes \mathbf{B})=\rho\mathbf{F},\\
	\label{eq:energy}
	&\frac{\partial E}{\partial t}+\nabla\cdot ((E+P^*)\mathbf{v}-(\mathbf{B}\cdot\mathbf{v})\mathbf{B})=\rho\mathbf{F}\cdot\mathbf{v}+Q-\mathcal{L},\\
	\label{eq:induction}
	&\frac{\partial\mathbf{B}}{\partial t}-\nabla\times(\mathbf{v}\times\mathbf{B})=0,\\
	\label{eq:pressure}
	&P^*=P+\frac{\mathbf{B}\cdot\mathbf{B}}{2},\\
	\label{eq:tot_energy}
	&E=\frac{\rho\mathbf{v}\cdot\mathbf{v}}{2} + \frac{P}{\gamma-1}+\frac{\mathbf{B}\cdot\mathbf{B}}{2}.
	\end{align}
\end{subequations}

Here, $\rho$ is the gas mass density, $\mathbf{v}$ is the velocity field, $\mathbf{B}$ is the magnetic field, and $P = \rho k_B T / (\mu m_p)$ is the thermal pressure. The term $\mathbf{F}$ represents the turbulent acceleration field, and $E$ is the total energy density. Additional constants include the mean molecular weight $\mu=0.6000317$ \footnote{Note that we fix $\mu$ to a constant value at all temperatures and have not considered the effect of its value changing at $T\lesssim10^4~\mathrm{K}$, which is expected to decrease our number density and cooling rate estimates at these temperatures. But since the cooling rate drops substantially below $10^4~\mathrm{K}$, we do not expect this to have a strong effect on the outcome of our simulations.}, proton mass $m_p$, Boltzmann constant $k_B$, and temperature $T$. The heating and cooling rate densities are denoted by $Q(t)$ and $\mathcal{L}(\rho,T)$, respectively, and we adopt an adiabatic index $\gamma = 5/3$.

\subsection{Cooling, Heating, and Turbulence Forcing}\label{subsec:heating_cooling_implementation}

\begin{subequations}
The radiative cooling function is defined as:
\begin{equation}
    \mathcal{L}(\rho,T) = n_H^2\Lambda(T), \label{eq:cooling_function}
\end{equation}
where $\Lambda(T)$ is a temperature-dependent photo+collisional ionization equilibrium cooling function generated using \texttt{AstroPlasma}\footnote{\url{https://github.com/dutta-alankar/AstroPlasma}} for a metallicity of $0.3Z_\odot$, in the presence of \cite{HaardtMadau2012ApJ} ionizing background radiation. The hydrogen number density is given by $n_H = \rho x_H / m_H$, with $x_H = 0.715$ and $m_H$ the hydrogen mass. We impose a temperature floor of $10^{3.2}~\mathrm{K}$.

To maintain global thermal balance and prevent runaway cooling, we introduce a compensatory heating term. This term offsets the net energy loss due to radiative cooling and turbulent forcing, and is defined as:
\begin{equation}
    Q = \frac{\rho^{\alpha_\mathrm{heat}}\max\left[\left(\int (\mathcal{L}-\rho \mathbf{F}\cdot\mathbf{v})\mathrm{d}V\right),0\right]}{\int\rho^{\alpha_\mathrm{heat}}\mathrm{d}V}, \label{eq:Q_heat}
\end{equation}
where $\alpha_\mathrm{heat} = 0$ corresponds to volume-weighted heating and $\alpha_\mathrm{heat} = 1$ to mass-weighted heating.
\end{subequations}

We model the turbulent acceleration field $\mathbf{F}$ using a spectral forcing method based on the stochastic Ornstein—Uhlenbeck (OU) process \citep{eswaran1988examination,schmidt2006numerical,federrath2010}. This method allows us to impose a finite autocorrelation timescale, set as $t_{\mathrm{corr}} = \ell_\mathrm{driv}/v_\mathrm{driv}$ across all simulations, where $\ell_\mathrm{driv}$ is the turbulence driving scale and $v_\mathrm{driv}$ is the target rms velocity. Turbulence is injected at large scales, with modes satisfying $1 \leq |\mathbf{k}|L/2\pi \leq 3$. The forcing power spectrum follows a parabolic profile, peaking at $k_\mathrm{peak} = 2$ (expressed in units of $2\pi/L$ for simplicity). At smaller scales ($k > 3$), turbulence develops self-consistently.

We control the nature of the forcing—solenoidal or compressive—by projecting the acceleration field onto components perpendicular or parallel to $\mathbf{k}$, respectively. The amplitude of the forcing is dynamically scaled at each timestep to match a specified energy injection rate $\dot{E}_\mathrm{turb}$. Compressively forced turbulence typically requires a larger energy injection rate to achieve a similar velocity dispersion, since a fraction of the energy does not cascade down from large scales to intermediate scales all the way to the dissipation scale. Rather, it is dissipated instantaneously through shocks, even when the rms velocity is subsonic. For further details, see Section 2.1 of \citet{federrath2010}.

\subsection{Numerical Methods}\label{subsec:numerical_methods}

We evolve equations \crefrange{eq:continuity}{eq:tot_energy} using \texttt{AthenaK}\footnote{\url{https://github.com/IAS-Astrophysics/athenak}}, a GPU-enabled, performance-portable version of \texttt{Athena++} \citep{Stone2020ApJS,Stone2024arXiv}, built on the Kokkos library \citep{Trott2021CSE}. Our numerical scheme employs second-order Runge–Kutta (RK3) time integration, the HLLD Riemann solver\footnote{For purely hydrodynamic simulations, we use the HLLC solver.}, and piecewise parabolic spatial reconstruction. To handle unphysical values in velocity or temperature, we apply a first-order flux correction algorithm as described in \citet{Lemaster2009ApJ}.

\subsection{Key time and length scales}\label{subsec:theory_time_length_scales}

Associated with the source terms due to cooling and turbulence driving defined above are some key length and time scales relevant to our simulations. 
The turbulence mixing time at a scale $\ell$ and the cooling time (scale independent) are given by:
\begin{subequations}
    \begin{align}
        t_\mathrm{mix}^\ell &= \ell/v_\ell\text{, and} \label{eq:t_mix}\\
        t_\mathrm{cool}&=\frac{nk_BT}{(\gamma-1)n_H^2\Lambda(T)} \label{eq:t_cool},
    \end{align}
    respectively. For the remainder of the paper, we shall refer to $t_\mathrm{mix}^{\ell_\mathrm{int}}$ as $t_\mathrm{mix}$, where the integral scale $\ell_\mathrm{int}$ is defined below in \cref{eq:integral_length}. 
    For multiphase gas, the mixing time between the different temperature phases has a further dependence on the density contrast $\chi=\rho_\mathrm{cold}/\rho_\mathrm{hot}$, where 
    \begin{equation}
        t_\mathrm{mix}^\mathrm{multi} = \sqrt{\chi}t_\mathrm{mix} \label{eq:t_mix_multi}.
    \end{equation}
The turbulence driving scale, the integral scale, the cooling length, and the Turbulent Field's length are given by the following equations:

    \begin{align}
        \ell_\mathrm{driv} &= \frac{1}{k_\mathrm{peak}}, \label{eq:l_driv}\\
        \ell_\mathrm{int} &= \frac{\int k^{-1} E(k)\,\mathrm{d}k}{\int E(k)\,\mathrm{d}k}, \label{eq:integral_length}\\
        \ell_\mathrm{cool}(n,T) &= c_s t_\mathrm{cool}, \label{eq:cooling_length}\\
        \lambda_{\rm F,turb} &\equiv 
        \max_{\ell}\left\{\, \ell \;\big|\; t_{\rm mix}(\ell) < t_{\rm cool} \right\}.\label{eq:Lambda_F_turb}
    \end{align}
The integral scale is the energy-containing scale of turbulence and is slightly smaller than the turbulence driving scale in our simulations.

\end{subequations}

\subsection{Initial Conditions}\label{subsec:init_conditions}

We conduct a suite of 19 simulations to explore the parameter space of multiphase MHD turbulence in the CGM at $0.125$--$8$ kpc and its impact on the medium's thermodynamic and kinematic properties. A summary of the simulation parameters is provided in \Cref{tab:sim_params}.

In all simulations—except for the compressively driven \texttt{CompHydroLR} run—we apply solenoidal (divergence-free) turbulence, injected at a driving scale of $\ell_\mathrm{driv} = \ell_\mathrm{box}/2$. Motivated by observations (see \cref{fig:vsf_L1}), for our \texttt{Fiducial} set of simulations with a box size of $1~\mathrm{kpc}$, we target a velocity dispersion of approximately $20~\mathrm{km/s}$. The initial temperature is set to $10^6~\mathrm{K}$, the expected virial temperature of a Milky Way-mass halo, and the number density is $3\times10^{-3}~\mathrm{cm}^{-3}$, to represent a dense cloud complex with abundant multiphase gas within the halo. To study lower-density regions of the CGM, we use the \texttt{LDens} set, which has a lower density of $3\times10^{-4}~\mathrm{cm}^{-3}$. To test the effect of the density contrast $\chi$ on our results, we conduct an additional \texttt{ICM} set of three simulations where the initial density is $0.3~\mathrm{cm}^{-3}$ and the initial temperature is $10^7~\mathrm{K}$, mimicking typical ICM conditions in the central regions of a cool-core cluster.

All simulations employ a density-dependent heating function ($\alpha_\mathrm{heat} = 1$). To aid in the formation of multiphase gas and to crudely mimic realistic CGM with satellites and filaments, in all our simulations we seed large-scale density fluctuations between $k=1$—$3$, with $\sigma_\rho/\mean{\rho}\approx0.6$. Large perturbations are necessary to seed cold filaments in our setup, we refer the reader to \cite{Mohapatra2023MNRAS,Mohapatra2024ApJ} for further details on how  multiphase gas formation depends on simulation properties such as the ratio $t_{\rm cool}/t_{\rm mix}$.

Our initial conditions are guided by the cloud density-size and mass-size relations presented in \cite{HWChen2023ApJ}. Among the suite of simulations, the \texttt{LDens} runs most closely replicate the typical CGM environments inferred from photoionization modeling of absorption-line data. However, as we demonstrate below, these runs fail to sustain a multiphase CGM due to rapid mixing at scales $\lesssim 1~\mathrm{kpc}$. Furthermore, absorption-line observations tend to be more sensitive to extended, lower-density gas along the line-of-sight, in contrast to the higher-density regions that dominate emission signatures. To account for this, our  \texttt{Fiducial} runs adopt slightly denser initial conditions to increase the probability of cold gas formation and survival. These setups can be interpreted as denser turbulent cloud complexes embedded within a lower-density CGM background, akin to those described in \citet{Bisht2025MNRAS}, encompassing a range of masses and sizes in the cold phase.

\subsubsection{Magnetic Fields}\label{subsubsec:mag_fields_init}

All MHD simulations are initialized with a random magnetic field corresponding to a plasma beta $(\beta) = 100$, and no imposed mean-field component. This setup is broadly consistent with magnetic field strengths of $\sim 0.1~\mu\mathrm{G}$ reported in the CGM of nearby galaxies \citep{Heesen2023A&A}. The Fourier amplitude of the initial magnetic field, $B_k$, follows a power-law distribution in wavenumber space between $k = 4$ and $k = 12$, with a slope of $k^{-1/3}$, reflecting a Kolmogorov-like scaling. We note that this slope is somewhat shallower than the $k^{-9/5}$ magnetic power spectrum scaling (ours is $k^{-5/3}$) observed in high-resolution MHD simulations of multiphase turbulence \citep{Fielding2023ApJ}. An exception to this setup is the \texttt{MHDUni} run, which employs a uniform magnetic field with the same plasma beta, but without any random component.

\subsubsection{Varying the Box Size} \label{subsec:vary_box_size_methods}

To contextualize our results within large-volume cosmological simulations, we investigate the effects of varying the simulation box size. Our  \texttt{Fiducial} setup employs a box of size $1~\mathrm{kpc}$ with a resolution of $1024^3$ grid cells. In addition, we perform simulations with box sizes $\ell_{\mathrm{box}} = 8~\mathrm{kpc}$ and $0.125~\mathrm{kpc}$.

We adopt two distinct approaches to scale turbulence across these simulations:

    In the first approach, we maintain a constant energy injection rate per unit volume while varying the box size, representing a steady turbulent cascade to smaller scales. Under Kolmogorov scaling, the driving velocity and mixing time at the integral scale as:
    \begin{subequations}
    \begin{align}
        u_{\ell_\mathrm{box}} &= u_\texttt{fid} \left( \frac{\ell_\mathrm{box}}{1~\mathrm{kpc}} \right)^{1/3}, \label{eq:dedt_sig_v} \\
        t_\mathrm{mix} &= t_{\mathrm{mix},\texttt{fid}} \left( \frac{\ell_\mathrm{box}}{1~\mathrm{kpc}} \right)^{2/3}, \label{eq:dedt_tmix}
    \end{align}
    \end{subequations}
    where $u_{\ell_\mathrm{box}}$ is the driving velocity at scale $\ell_\mathrm{box}$ and $t_\mathrm{mix}$ is the corresponding mixing time. As the box size decreases, the mixing time shortens. Simulations following this scaling are labeled with \texttt{Dedt}.

    In the second approach, we fix the mixing time $t_\mathrm{mix}$ across all box sizes. The corresponding scaling relations are:
    \begin{subequations}
    \begin{align}
        t_\mathrm{mix} &= t_{\mathrm{mix},\texttt{fid}}, \label{eq:t_mix_fixed_tmix} \\
        u_{\ell_\mathrm{box}} &= u_\texttt{fid} \left( \frac{\ell_\mathrm{box}}{1~\mathrm{kpc}} \right), \label{eq:sigma_v_fixed_tmix}
    \end{align}
    \end{subequations}
    In this case, the driving velocity decreases with decreasing box size, resulting in weaker turbulence in smaller boxes. These simulations are labeled with \texttt{Tmix}.

\subsubsection{Varying the density contrast} \label{subsec:vary_dens_contrast}
We explore the impact of varying the density contrast $\chi$ between the hot and cold phases on the mixing rate between them, using our \texttt{ICM} suite of three simulations. Each simulation is initialized with a temperature $T = 10^7~\mathrm{K}$ and number density $n = 0.3~\mathrm{cm}^{-3}$. Given that the hot phase is at $10^7~\mathrm{K}$ and the cold phase is at approximately $10^4~\mathrm{K}$, these runs correspond to a density contrast of $\chi \approx 1000$, which is an order of magnitude larger than the CGM-like simulations discussed elsewhere in this paper.
All simulations use a box size of $0.125~\mathrm{kpc}$ and are directly compared to the \texttt{0.125TmixHydroLR} reference run. We conduct three sets of simulations:

 In the first run (\texttt{0.125ICMTmixHydroLR}), we maintain a fixed ratio of $t_\mathrm{cool}^{\mathrm{int}} / t_\mathrm{mix}$ relative to the \texttt{0.125TmixHydroLR} run. Here, $t_\mathrm{cool}^{\mathrm{int}}$ denotes the cooling time evaluated at the geometric mean temperature $T = \sqrt{T_\mathrm{cold} T_\mathrm{hot}}$, and number density $n=\sqrt{n_\mathrm{cold}n_\mathrm{hot}}$, and $t_\mathrm{mix}$ is the mixing time at the integral scale, see \cref{eq:t_mix}.

 In the second run (\texttt{0.125ICMTmixMPHydroLR}), we fix the ratio $t_\mathrm{cool}^{\mathrm{int}} / t_\mathrm{mix}^{\mathrm{multi}}$ relative to the reference run, where $t_\mathrm{mix}^{\mathrm{multi}}$ is defined in \cref{eq:t_mix_multi}. Since $t_\mathrm{mix}^{\mathrm{multi}}$ is longer for a larger $\chi$, these simulations require stronger turbulent driving to achieve comparable mixing.

In the third run (\texttt{0.125ICMTmixMPRestartHydroLR}), we initialize the simulation with conditions identical to \texttt{0.125TmixHydroLR}. Once a steady state is reached, we switch to a stronger turbulent driving regime, similar to that used in \texttt{0.125ICMTmixMPHydroLR}.

Finally, we test the effects of our assumptions such as using density versus volume-weighted heating (by changing $\alpha_{\rm heat}$), and switching off global thermal balance in \S\ref{sec:appendix_diff_heating_effect}. We test our results for convergence with resolution in \S\ref{sec:appendix_resolution_effect}.

\begin{table*}[!p]
\centering
\makebox[\textwidth][c]{%
  \rotatebox{90}{%
    \begin{minipage}{\textheight}
      \centering
      \caption{Simulation parameters and statistics for different runs}
      \label{tab:sim_params}
      \resizebox{\textheight}{!}{%
        \renewcommand{\arraystretch}{1.5}
        \begin{tabular}{lcccccccccr}
        \hline\hline
        Label & Resolution & $\ell_\mathrm{box}$ (kpc) & $n$ (cm$^{-3}$) & $\dot{E}_\mathrm{turb}$ (10$^{36}$\,erg\,s$^{-1}$) & $t_\mathrm{cool}/t_\mathrm{mix}$ & $\sigma_v$ (km\,s$^{-1}$) & $M_\mathrm{cold}/M_\mathrm{tot}$ & $V_\mathrm{cold}/V_\mathrm{tot}$ & $A_\mathrm{int}/A_\mathrm{tot}$ & $t_\mathrm{surv}$ (Myr) \\
        {\scriptsize(1)} & {\scriptsize(2)} & {\scriptsize(3)} & {\scriptsize(4)} & {\scriptsize(5)} & {\scriptsize(6)} & {\scriptsize(7)} & {\scriptsize(8)} & {\scriptsize(9)} & {\scriptsize(10)} & {\scriptsize(11)} \\
        \hline\hline
        \texttt{FidHydro}         & $1024^3$ & $1$     & $3\times10^{-3}$  & $0.138$               & $3.2$ & $19$  & $0.20$        & $3\times10^{-3}$  & $0.83$ & $>200$ \\
        \texttt{FidMHD}           & $1024^3$ & $1$     & $3\times10^{-3}$  & $0.138$               & $2.5$ & $15$  & $0.50$        & $1\times10^{-2}$  & $0.84$ & $>200$ \\
        \texttt{MHDUni}           & $1024^3$ & $1$     & $3\times10^{-3}$  & $0.138$               & $3.2$ & $19$  & $0.40$        & $8\times10^{-3}$  & $0.79$ & $>200$ \\ \hline
        \texttt{LDensHydro}       & $1024^3$ & $1$     & $3\times10^{-4}$  & $0.0138$              & $35$  & $21$  & $0.00$        & $0.00$            & $0.00$ & $50$ \\
        \texttt{LDensMHD}         & $1024^3$ & $1$     & $3\times10^{-4}$  & $0.0138$              & $27$  & $16$  & $10^{-3}$     & $2\times10^{-5}$  & $0.12$ & $150$ \\ \hline
        \texttt{8DedtHydroLR}     & $512^3$  & $8$     & $3\times10^{-3}$  & $70.7$                & $0.8$ & $40$  & $0.50$        & $1\times10^{-2}$  & $0.71$ & $>200$ \\
        \texttt{8DedtMHDLR}       & $512^3$  & $8$     & $3\times10^{-3}$  & $70.7$                & $0.7$ & $32$  & $0.50$        & $7\times10^{-3}$  & $0.70$ & $>200$ \\
        \texttt{8TmixHydroLR}     & $512^3$  & $8$     & $3\times10^{-3}$  & $4522$                & $3.1$ & $147$ & $0.14$        & $8\times10^{-3}$  & $0.92$ & $>200$ \\
        \texttt{8TmixMHDLR}       & $512^3$  & $8$     & $3\times10^{-3}$  & $4522$                & $2.9$ & $140$ & $0.18$        & $0.10$            & $0.93$ & $>200$ \\
        \texttt{0.125DedtHydroLR} & $512^3$  & $0.125$ & $3\times10^{-3}$  & $2.7\times10^{-4}$    & $13.5$& $10$  & $0.00$        & $0.00$            & $0.00$ & $35$ \\
        \texttt{0.125DedtMHDLR}   & $512^3$  & $0.125$ & $3\times10^{-3}$  & $2.7\times10^{-4}$    & $10.1$& $7.5$ & $0.01$        & $2\times10^{-4}$  & $0.29$ & $>200$ \\
        \texttt{0.125TmixHydroLR} & $512^3$  & $0.125$ & $3\times10^{-3}$  & $4.2\times10^{-6}$    & $1.9$ & $1.4$ & $0.30$        & $5\times10^{-3}$  & $0.64$ & $>200$ \\
        \texttt{0.125TmixMHDLR}   & $512^3$  & $0.125$ & $3\times10^{-3}$  & $4.2\times10^{-6}$    & $2.2$ & $1.6$ & $0.40$        & $4\times10^{-3}$  & $0.27$ & $>200$ \\ \hline
        \texttt{CompHydroLR}      & $512^3$  & $1$     & $3\times10^{-3}$  & $0.55$                & $2.2$ & $13$  & $0.34$        & $6\times10^{-3}$  & $0.40$ & $>200$ \\ \hline
        \texttt{FidHydroLR}       & $512^3$  & $1$     & $3\times10^{-3}$  & $0.138$               & $3.0$ & $18$  & $0.20$        & $3\times10^{-3}$  & $0.62$ & $>200$ \\
        \texttt{FidMHDLR}         & $512^3$  & $1$     & $3\times10^{-3}$  & $0.138$               & $2.7$ & $16$  & $0.42$        & $8\times10^{-3}$  & $0.59$ & $>200$ \\ \hline
        \texttt{0.125ICMTmixHydroLR}  & $512^3$ & $0.125$ & $3\times10^{-2}$ & $1.38\times10^{-4}$ & $2.0$ & $8$  & $0.50$        & $1\times10^{-3}$  & $0.25$ & $>140$ \\
        \texttt{0.125ICMTmixMPHydroLR}& $512^3$ & $0.125$ & $3\times10^{-2}$ & $0.043$            & $6.25$ & $25$ & $0.00$        & $0.00$            & $0.00$ & $0$ \\
        \texttt{0.125ICMTmixMPRestartHydroLR}& $512^3$ & $0.125$ & $3\times10^{-2}$ & $0.043$     & $6.25$ & $25$ & $0.00$        & $0.00$            & $0.00$ & $20$ \\ 
        \hline \hline
        \end{tabular}
      }
      \par\medskip
      \justifying{\footnotesize Notes: Column (1) lists the simulation labels; Column (2) shows the resolution; Column (3) gives the box size. Column (4) is the turbulence energy injection rate (in units given in the header). Column (5) denotes the initial ratio of cooling to mixing time at the driving scale. Columns (6)--(9) report the velocity dispersion, the average mass and volume fractions of cold gas, and the area covering fraction of cold and intermediate-temperature gas ($T\leq10^{5}\,\mathrm{K}$, $N>10^{18}\,\mathrm{cm}^{-2}$) over the final $100\,$Myr, respectively. Column (10) indicates the survival time of cold gas, defined as the time after which its mass fraction drops below $10^{-4}$.}
    \end{minipage}%
  }%
}
\end{table*}

\section{Results}\label{sec:results}

\subsection{Morphology and Emission Structure}\label{subsec:morphology_emission}
\begin{figure*}
		\centering
	\includegraphics[width=2.0\columnwidth]{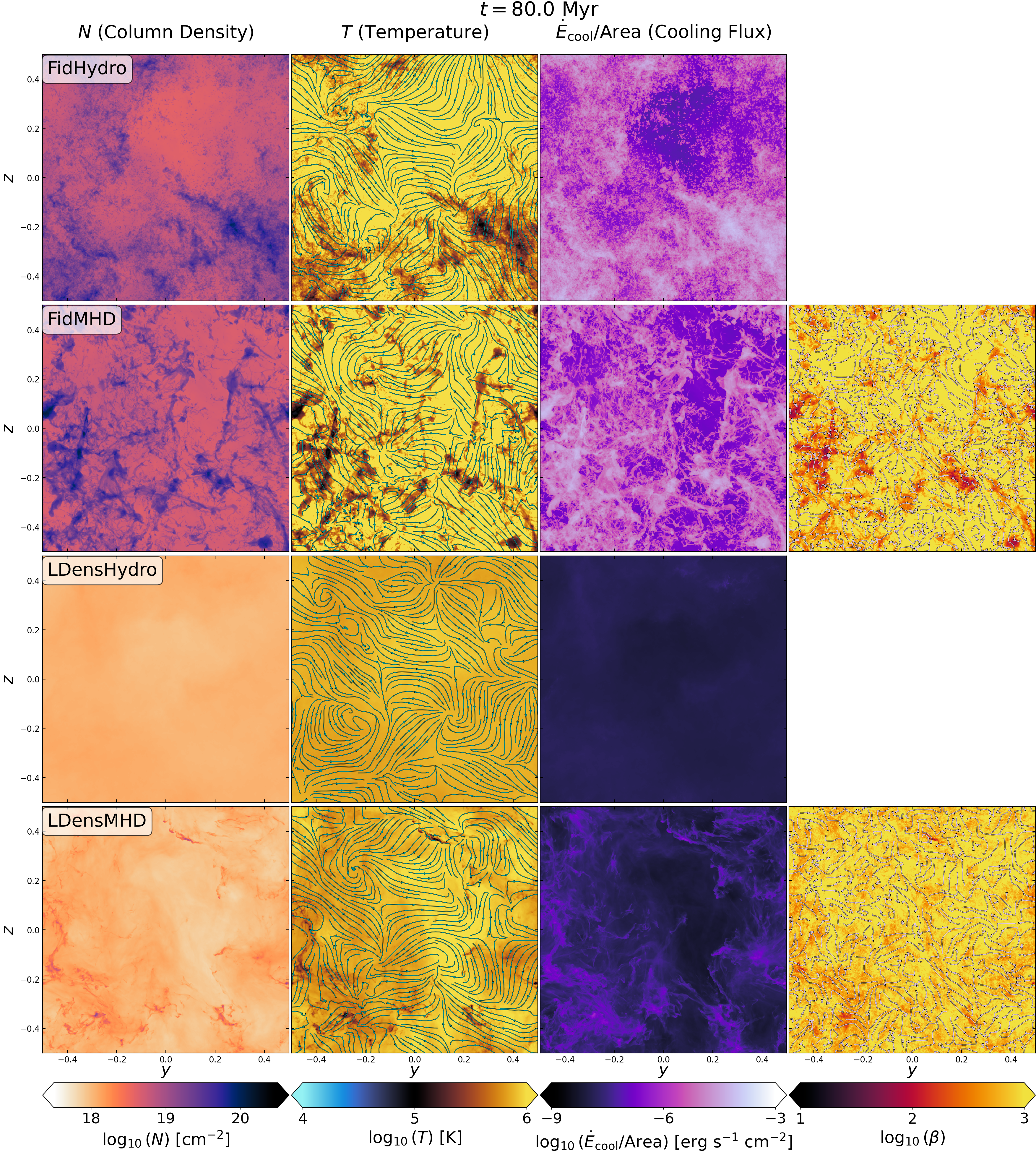}	
	\caption{Projections of density (Col 1, vol-weighted), temperature (Col 2, mass-weighted), Cooling rate (Col 3, vol-weighted), and plasma beta (Col 4, mass-weighted) along the $x$-direction for the  \texttt{Fiducial} and lower density \texttt{LDens} sets of runs. The streamlines on Columns 2 and 4 depict the mass-weighted projections of the velocity and magnetic field, respectively. Cold, dense gas exists in all runs except the \texttt{LDensHydro} run, where it has mostly evaporated by $t=80~\mathrm{Myr}$. The \texttt{FidHydro} run shows a lot of small, shattered cold clouds that are absent in the FidMHD run. In the MHD runs, we find the cold clouds are filamentary and have a larger fraction of magnetic pressure.  An animated version of this figure is available \href{https://youtu.be/mz5G-TjZSkU?si=EWulrtROUvNhikxa}{here}, as well as in the arXiv submission.}
	\label{fig:projection}
\end{figure*}
\begin{figure*}
		\centering
	\includegraphics[width=2.0\columnwidth]{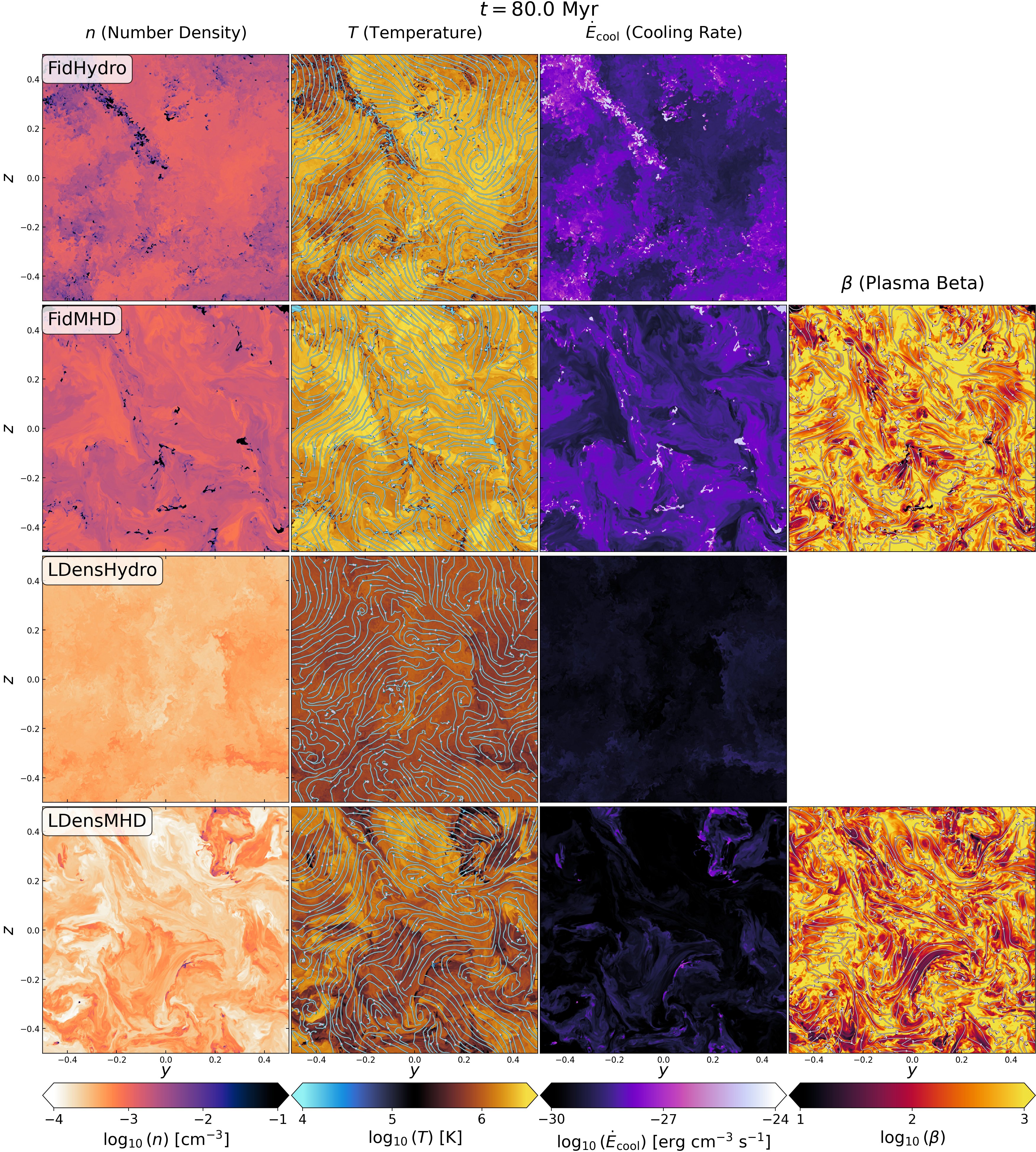}	
	\caption{Similar to \Cref{fig:projection}, except here we show the slices of density (Col 1), temperature (Col 2), Cooling rate (Col 3), and plasma beta (Col 4) along the $x$-direction for the  \texttt{Fiducial} and lower density \texttt{LDens} sets of runs at $t=80~\mathrm{Myr}$. The streamlines on Columns 2 and 4 depict the velocity and magnetic field, respectively. We find more gas at intermediate temperatures/densities for the \texttt{FidHydro} run, compared to the \texttt{FidMHD} run, due to suppressed mixing in the MHD run. An animated version of this figure is available \href{https://www.youtube.com/shorts/PAFFfJaBDCw}{here}.}
	\label{fig:slices}
\end{figure*}

In \Cref{fig:projection}, we show projections along the $x$-direction at $t = 80~\mathrm{Myr}$ for four key quantities: gas column density (\emph{Col 1}), mass-weighted temperature (\emph{Col 2}), net emission ($\int n_H^2\Lambda(T)\,\mathrm{d}x$, \emph{Col 3}), and mass-weighted plasma beta (\emph{Col 4}). These are presented for both the  \texttt{Fiducial} (\texttt{FidHydro}, \texttt{FidMHD}) and lower-density (\texttt{LDensHydro}, \texttt{LDensMHD}) simulation sets. Corresponding slices in the $yz$-plane are shown in \Cref{fig:slices}. An animated version of \Cref{fig:slices} is available in the online version of this manuscript and on \href{https://www.youtube.com/shorts/PAFFfJaBDCw}{YouTube}.

All simulations develop a multiphase medium, but the \texttt{LDensHydro} run (third row) shows a rapid decline in cold gas mass fraction, falling below $10^{-4}$ by $t = 60~\mathrm{Myr}$. At this stage, remnants of the cold phase—now heated to $T \sim 10^5~\mathrm{K}$—are visible as they mix into the hot medium, eventually leading to a single-phase state. In contrast, the cold gas in the other runs remains dense, and in the MHD cases, exhibits low plasma beta values. Since net emission scales as $n_H^2\Lambda(T)$ and peaks near $10^4~\mathrm{K}$, the emission maps closely trace the distribution of cold, dense gas. 
Even though the cold gas dominates net emission, it does not dominate the mass along any line-of-sight (LOS), since the mass-weighted average temperature is around $\approx10^5$--$10^6~\mathrm{K}$ along different LOS.

In the  \texttt{Fiducial} simulations, the hydrodynamic run produces numerous small cold clouds, formed via fragmentation of large-scale filaments ($\sim100~\mathrm{pc}$) that arise from initial density perturbations. In contrast, the MHD run retains larger filamentary structures, spanning several hundred parsecs, supported by magnetic pressure during rapid collapse. This behavior is consistent with the predictions of \citet{Wang2025arXiv}, who argue that magnetic fields can suppress fragmentation during compressive cooling by providing non-thermal pressure support. As cold gas condenses, flux freezing amplifies the magnetic field, further enhancing this support.

The density and temperature slices reveal distinct structural differences between the hydro and MHD simulations. In the hydro run, cold gas is embedded within a broader envelope of intermediate-temperature material, resulting in relatively diffuse emission. In contrast, the \texttt{FidMHD} run (second row) exhibits emission that is strongly concentrated around cold gas regions. This contrast arises from the suppression of turbulent mixing due to magnetic fields. As shown in the rightmost panels of \Cref{fig:slices}, the white streamlines in the plasma beta slices trace the $yz$-plane magnetic field. Regions of low plasma beta are associated with coherent magnetic field structures that align with the cooler, denser gas. These magnetic fields inhibit mixing, thereby diminishing emission and absorption from intermediate-ionization species such as \texttt{SiIV} and \texttt{CIV}.

In the \texttt{LDens} runs (bottom two rows), both the cold gas mass and emission are significantly lower than in their denser \texttt{Fiducial} counterparts. This is primarily due to longer cooling times at lower densities, which limit the fraction of gas with sufficiently small $t_\mathrm{cool}/t_\mathrm{mix}$ to survive interaction with the hot phase. Nevertheless, in the \texttt{LDensMHD} run, magnetic fields suppress mixing, allowing cold gas to persist at $t = 80~\mathrm{Myr}$ and survive for much longer before being fully assimilated into the hot medium.

\subsection{Time evolution of Multiphase Structure}\label{subsec:time_evol_Fid_ldens}
\begin{figure*}
		\centering
	\includegraphics[width=2.0\columnwidth]{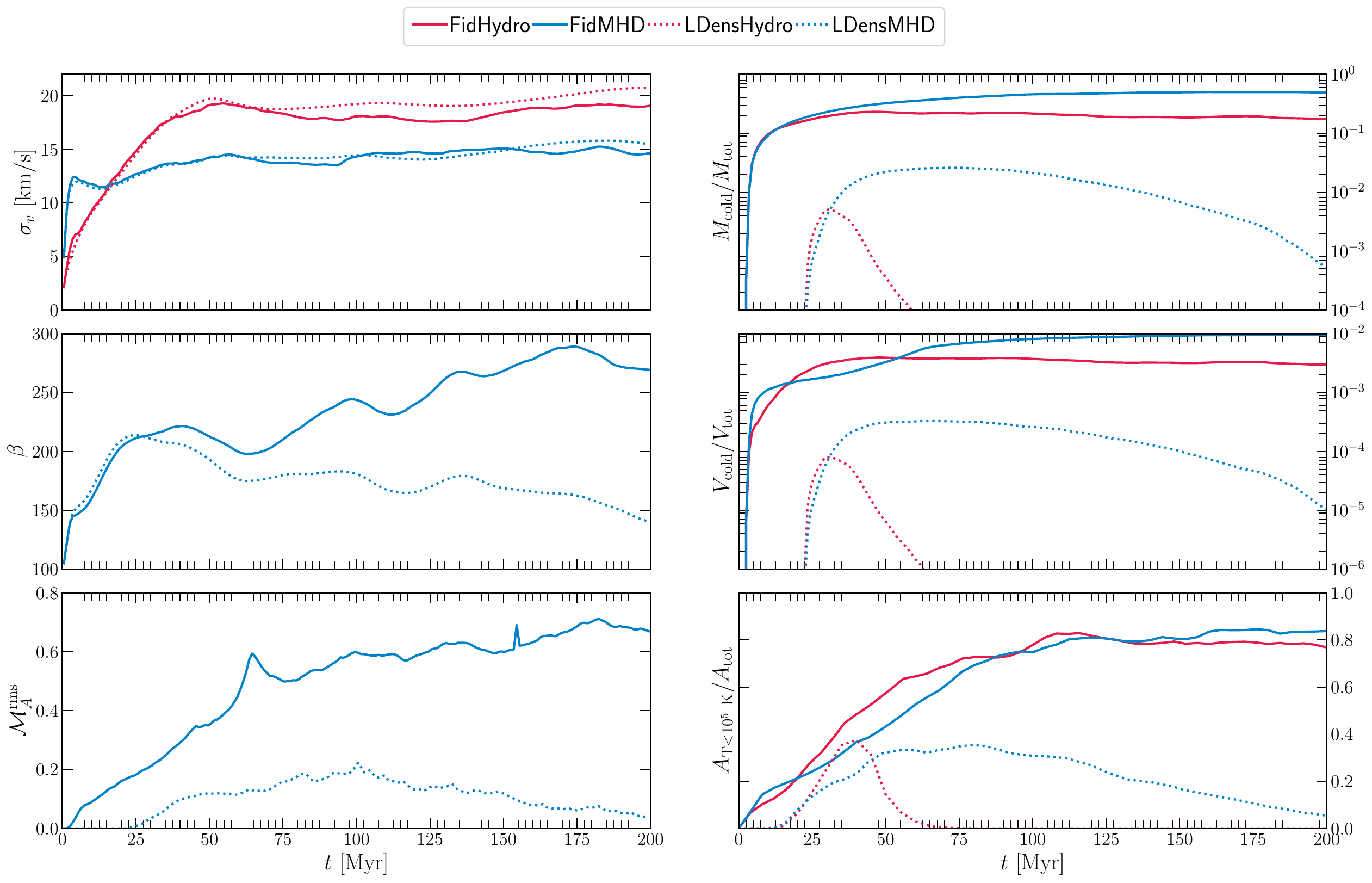}	
	\caption{Time evolution of different statistical properties of the gas for the \texttt{Fiducial} and lower density \texttt{LDens} hydro and MHD runs. \emph{Left col:} The velocity dispersion is smaller for the MHD runs compared to the hydro runs. For the MHD runs, the plasma beta is a few $100$ in steady state, and both simulations are sub-Alfvenic. \emph{Right col:} The \texttt{Fiducial} set of runs reach a steady state with roughly half of their mass in the cold phase ($T<10^{4.2}~\mathrm{K}$), which occupies $\lesssim1\%$ of the volume. For the \texttt{LDens} set of runs, $<1\%$ of the mass converts into the cold phase, and it quickly mixes up with the hot phase due to its longer cooling time. The Area covering fraction of gas at $T<10^5~\mathrm{K}$, with column density$>10^{18} \mathrm{cm}^{-2}$, is roughly $80\%$ for the \texttt{Fiducial} set, whereas it reaches a maximum of $40\%$ for the \texttt{LDens} set before dropping to $0$.}
	\label{fig:time_evol_fid_ldens}
\end{figure*}

In \Cref{fig:time_evol_fid_ldens}, we present the temporal evolution of key physical quantities for the  \texttt{Fiducial} and \texttt{LDens} simulation suites. The \texttt{Fiducial} runs are initialized with a gas density ten times higher than the \texttt{LDens} runs, and accordingly use a tenfold larger turbulent energy injection rate, $\dot{E}_\mathrm{turb}$, to maintain the same acceleration field. This energy injection rate is kept constant across both hydrodynamic and MHD simulations. As shown in the top-left panel, the MHD runs exhibit slightly lower values of velocity dispersion due to a portion of the injected energy being diverted into magnetic field reorganization. These velocities are consistent with the non-thermal motions inferred for $\sim1~\mathrm{kpc}$-scale CGM clouds in \citet{HWChen2023ApJ}.

The right column of \Cref{fig:time_evol_fid_ldens} shows the evolution of the mass and volume fractions of cold gas ($T < 10^{4.2}~\mathrm{K}$)\footnote{We choose $T\lesssim10^{4.2}~\mathrm{K}$ to define cold gas since it is slightly smaller than the temperature at which the cooling time reaches a minimum, as a function of temperature, assuming isobaric cooling. Small changes in this definition are not expected to affect our quantitative results, since any gas that cools to $10^{4.2}~\mathrm{K}$ will cool to further lower temperatures.}, as well as the area covering fraction of cold and intermediate-temperature gas ($T < 10^5~\mathrm{K}$, column density $N > 10^{18}~\mathrm{cm}^{-2}$). In the \texttt{LDens} runs 
the cold gas reaches a peak mass fraction of a few percent in the MHD case and less than $1\%$ in the hydro case. Its volume fraction remains below $\sim0.01\%$. In the \texttt{LDensHydro} run, the cold gas rapidly disappears within a few mixing timescales, while in the \texttt{LDensMHD} run, magnetic fields suppress mixing, allowing the cold gas to persist longer. However, it eventually declines below $10^{-4}$. 

The area covering fraction of gas with $T < 10^5~\mathrm{K}$ peaks at $\sim40\%$ in both \texttt{LDens} runs and decreases as the cold gas is assimilated into the hot phase. In contrast, the  \texttt{Fiducial} runs show significantly more robust cold gas formation, with steady-state mass fractions reaching $\sim20\%$ in hydro and up to $\sim50\%$ in MHD. Although the volume fraction remains $\lesssim1\%$, the area covering fraction reaches $\sim80\%$. These results suggest that while cold gas can form in low-density CGM environments (as modeled in the \texttt{LDens} runs), its survival is uncertain—even in the presence of magnetic fields. In denser CGM conditions, analogous to turbulent cloud complexes, \citep[for instance, as envisaged in ][]{Bisht2025MNRAS,Hummels2024ApJ} our results indicate that a steady-state multiphase structure can be sustained, with a substantial mass in the cold phase, that has a low volume filling, but a significantly high area covering fraction.

In our MHD simulations, we initialize a random magnetic field with no net flux and $\beta = 100$ (see \S\ref{subsubsec:mag_fields_init}). Both MHD runs remain sub-Alfv\'enic throughout the simulation, although the \texttt{FidMHD} run approaches trans-Alfv\'enic conditions by $t = 200~\mathrm{Myr}$. We observe that the volume-averaged $\beta$ increases over time for the \texttt{FidMHD} run. Since our simulations are maintained in global thermal balance (net cooling is balanced by turbulent energy injection and thermal heating, see \S\ref{subsec:init_conditions} and eq.~\ref{eq:Q_heat}), once cold gas forms and the net cooling rate increases, more thermal heat is pumped in, which increases the thermal pressure of the hot gas, leading to a larger $\beta$. For the \texttt{LDensMHD} run, additional thermal heating is weak or generally inactive due to its small cooling rate, so $\beta$ does not increase significantly during the simulation.

\subsection{Phase structure}\label{subsec:phase_structure}
\begin{figure*}
		\centering
	\includegraphics[width=2.0\columnwidth]{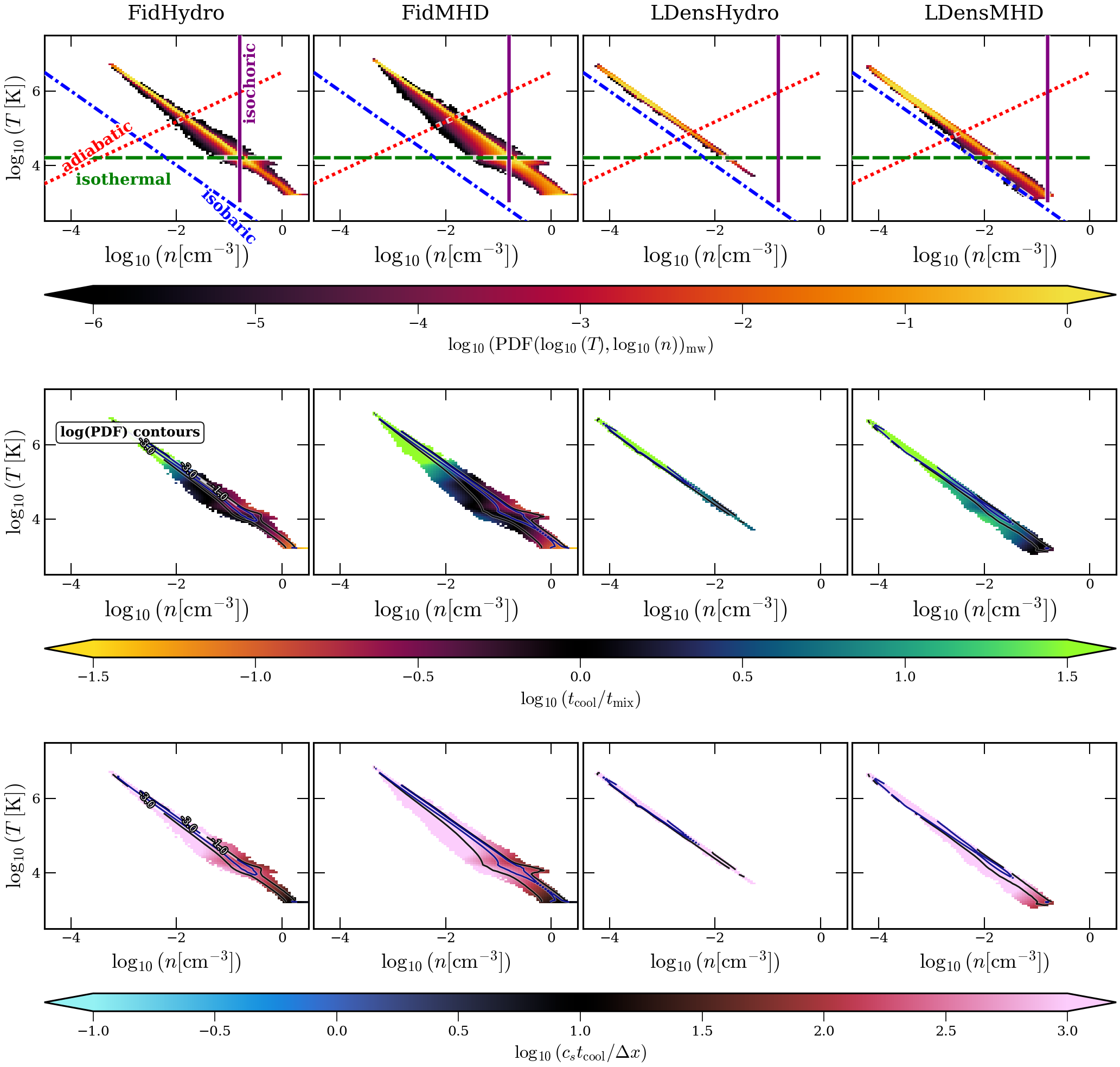}	
	\caption{Density-temperature phase diagram (\emph{Row 1}) for our \texttt{Fiducial} and lower density \texttt{LDens} hydro and MHD runs; also shown are $t_\mathrm{cool}/t_\mathrm{mix}$ and $c_st_\mathrm{cool}/\Delta x$ in \emph{Rows 2} and \emph{3}, respectively. These plots average over the phase diagrams from $t=0$ until $t=200~\mathrm{Myr}$. The contours on \emph{Rows 2} and \emph{3} denote the $90\%$ and $99.9\%$ percentile of the density-temperature PDF. The cold and hot gas are mostly isobaric. For \texttt{LDens} set of runs, $t_\mathrm{cool}>t_\mathrm{mix}$ for all gas, whereas for the \texttt{Fiducial} set, $t_\mathrm{cool}\lesssim t_\mathrm{mix}$ for $T\lesssim10^5~\mathrm{K}$. For gas at $T>10^4~\mathrm{K}$, we resolve $c_st_\mathrm{cool}$ at all temperatures and densities by at least 30 cells for all our simulations.}
	\label{fig:dens_temp_phase_diagram}
\end{figure*}

In this subsection, we examine the phase structure of the gas in our simulations and compare our findings against absorption-line photoionization modeling in observations \citep[e.g.][]{ZQu2022MNRASCUBS,ZQu2023MNRASCUBS}.

The top row of \Cref{fig:dens_temp_phase_diagram} shows the temperature-density phase diagrams for the  \texttt{Fiducial} and \texttt{LDens} simulation sets, averaged over the interval $t = 10$—$200~\mathrm{Myr}$. Across all runs, the hot and cold phases are broadly in pressure equilibrium. Although we set the cooling floor to $10^{3.2}~\mathrm{K}$, the cooling rate plummets below $10^4~\mathrm{K}$ and the gas reaches a quasi-stable phase at this temperature, evident in the isothermal spread in the phase diagram. This spread is most pronounced in the \texttt{FidMHD} run, where the cold phase shows a slightly wider distribution around the isobar, though it represents only a small fraction of the total gas. As shown earlier in \Cref{fig:projection,fig:slices}, $\beta\gtrsim 10$ in most regions, indicating that magnetic pressure does not dominate the overall pressure support.

Observationally, \citet{Zahedy2019MNRAS,ZQu2022MNRASCUBS,ZQu2023MNRASCUBS} infer the density, temperature, and thermal pressure of the CGM through photoionization modeling of absorption features in quasar spectra. \citet{ZQu2022MNRASCUBS} report that, within individual absorption systems—particularly those with kinematically aligned components—denser gas tends to exhibit higher thermal pressure (see their Figure 4). In our simulations, we observe only modest local variations in gas pressure at $T = 10^4~\mathrm{K}$, significantly smaller than the multi-order-of-magnitude fluctuations inferred from observations. This discrepancy can be reconciled if the cold gas in the CGM is supported predominantly by non-thermal pressure components, or if the observations are sampling different regions of the CGM with varying pressure but similar temperatures.

Furthermore, \citet{Zahedy2019MNRAS,ZQu2023MNRASCUBS} find that the maximum inferred density along each absorber sightline correlates with projected distance from the host galaxy in a manner consistent with the pressure profile of the hot CGM. These results support the interpretation that, on global scales, the cold and hot CGM phases are approximately in pressure equilibrium. This equilibrium is also evident in both our \texttt{Fiducial} and \texttt{LDens} simulation sets, despite their order-of-magnitude differences in average density and pressure.

The middle row of \Cref{fig:dens_temp_phase_diagram} displays the distribution of the ratio $t_\mathrm{cool}/t_\mathrm{mix}$, where the mixing time is computed at the integral scale of turbulence. This provides a conservative estimate, as smaller eddies have shorter mixing times, implying that the actual $t_\mathrm{cool}/t_\mathrm{mix}$ may be higher than shown. Contours marking the $90\%$ and $99.9\%$ percentiles of the phase diagram are overlaid on the color map. In the  \texttt{Fiducial} runs, gas with $T \gtrsim 10^{5.5}~\mathrm{K}$ has $t_\mathrm{cool}/t_\mathrm{mix} \gg 1$, while intermediate and cold gas typically has $t_\mathrm{cool}/t_\mathrm{mix} \lesssim 1$. In contrast, the \texttt{LDens} runs show $t_\mathrm{cool}/t_\mathrm{mix} \gtrsim 1$ across all temperatures and densities, explaining the inability of these runs to sustain cold gas for longer than a few $t_\mathrm{mix}$. Even rapidly cooling cold and intermediate-temperature gas is mixed into the hot phase before it can condense. In the  \texttt{Fiducial} runs, once cold gas forms, mixing with the hot phase produces intermediate-temperature gas with $t_\mathrm{cool}/t_\mathrm{mix} \lesssim 1$, which can further cool and regenerate cold gas. In steady state, this leads to a balance between cooling and mixing, leading to a long-term existence of the cold gas until the end of the simulation at $t=200~\mathrm{Myr}$. Our results are thus in agreement with \cite{Gronke2022MNRAS}, who argue that the cold gas survives if the cooling time of the mixed gas is shorter than the Kelvin-Helmholtz time of the cold gas clump.

The bottom row of \Cref{fig:dens_temp_phase_diagram} shows the ratio of the cooling length (defined in eq.~\ref{eq:cooling_length})  to the simulation resolution, with probability density contours overlaid as in the middle panel. Resolving the cooling length is a necessary condition for accurately resolving isobaric evolution. E.g., if all the gas in the simulation box cooled to a single grid cell, the density would rise by a factor $(\ell_\mathrm{box}/\Delta x)^3$, which, if less than the density contrast, will have a lower pressure. For gas at or above $10^4~\mathrm{K}$, we resolve the cooling length with at least 30 grid cells, and find that the hot and cold gas are roughly isobaric. \citet{Fielding2020ApJ,Abruzzo2024ApJ} found that pressure differences between hot and cold phases decrease in their TRML simulations when the minimum cooling length is well resolved. However, \cite{Wang2025arXiv,Sharma2025} argue that even when $c_s t_\mathrm{cool}$ is resolved across all temperatures, a pressure dip in the phase diagram persists. \cite{Sharma2025} attribute this dip to the compressive turbulent stress as it mixes with the cold phase. Compressed magnetic fields can also provide additional non-thermal pressure support to the cold gas, leading to further deviations from isobaric behavior. We observe a small pressure dip close to $10^4~\mathrm{K}$, which is more evident in the phase diagram of the \texttt{FidMHD} run, inline with their findings. 

In our previous studies of turbulence in the intracluster medium \citep{Mohapatra2022MNRASb,Mohapatra2023MNRAS}, we reported an isochoric drop near $T \sim 10^{5.5}~\mathrm{K}$, which was likely due to under-resolving $c_s t_\mathrm{cool}$ by factors of $100$—$1000$, given the higher gas densities and a spatial resolution of $\sim100~\mathrm{pc}$ in those simulations.

\subsection{Scale-dependent Statistics}\label{subsec:scale-dependence}
\begin{figure*}
		\centering
	\includegraphics[width=1.6\columnwidth]{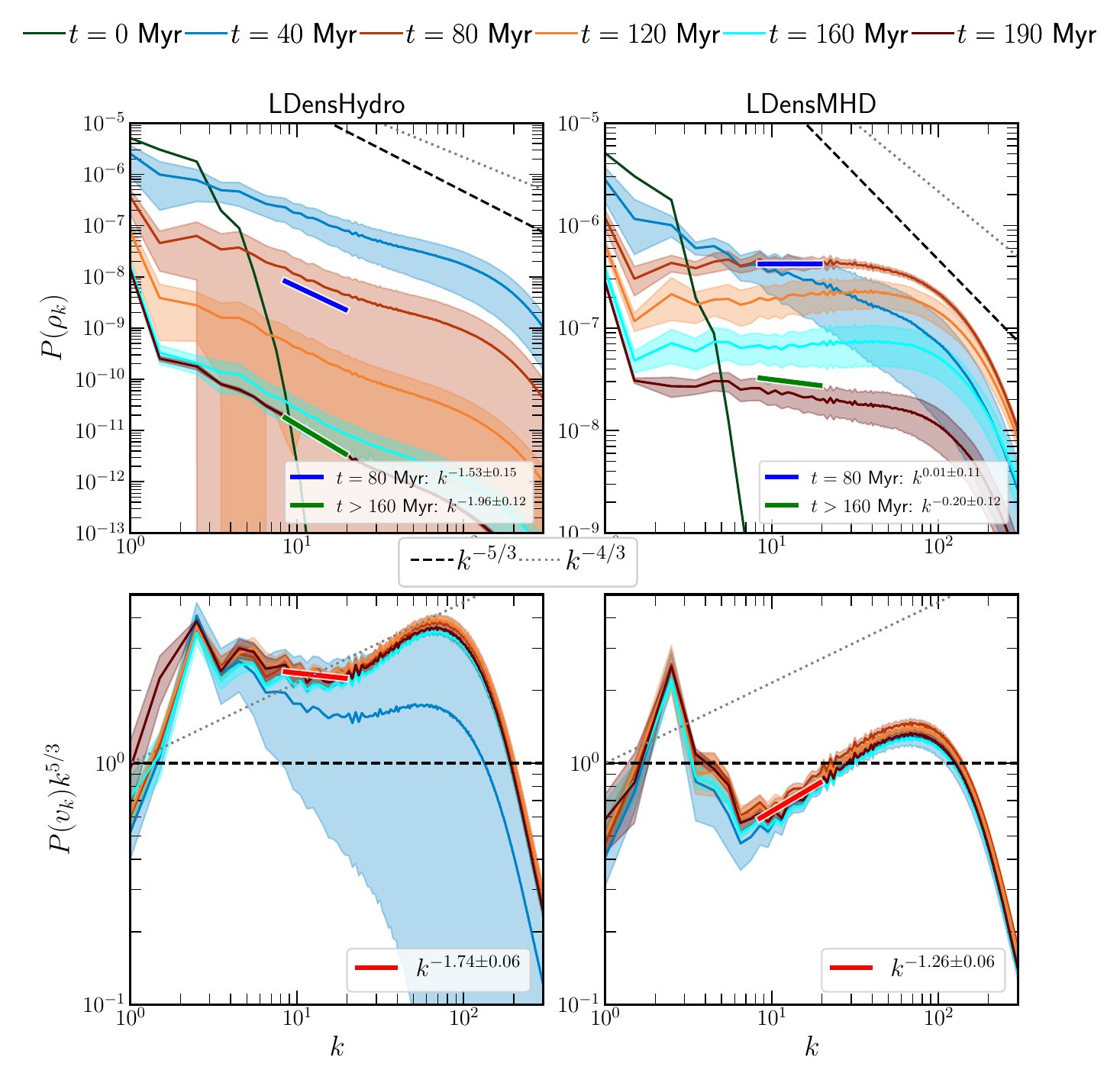}	
	\caption{Density and compensated velocity spectra at different times for our lower density \texttt{LDens} hydro and MHD runs. The density power spectrum is much flatter than Komlogorov (K41), and with time evolution it reduces in amplitude and becomes steeper for both hydro and MHD runs. The velocity power spectrum is slightly steeper than the K41 scaling for hydro runs, whereas for the MHD run it is slightly flatter than $k^{-4/3}$ scaling reported in recent multiphase turbulence studies.}
	\label{fig:spectra_LDens}
\end{figure*}

\begin{figure}
		\centering
	\includegraphics[width=\columnwidth]{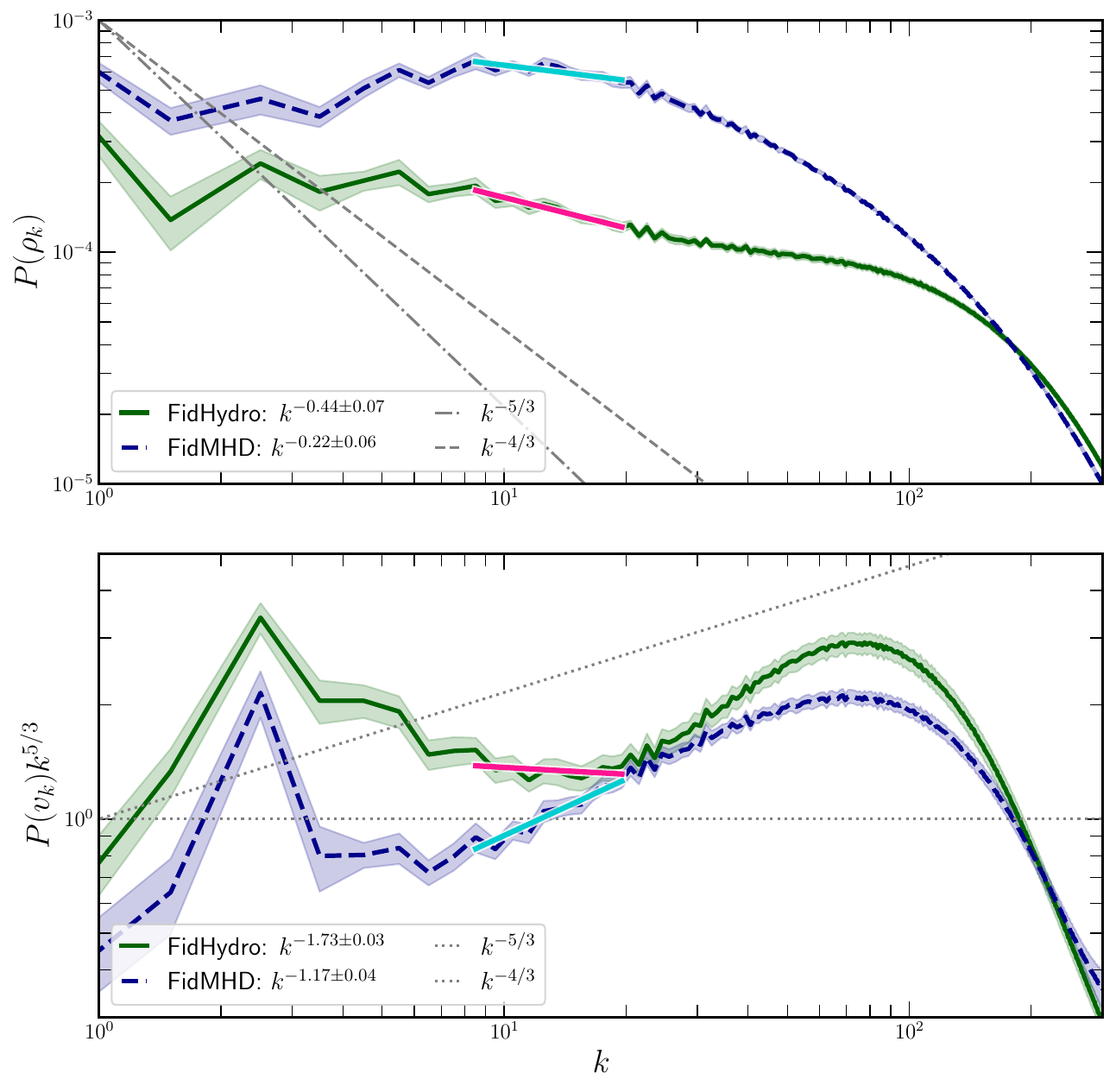}	
	\caption{Density and compensated velocity spectra in steady state for our \texttt{Fiducial} hydro and MHD runs. Similar to the \texttt{LDens} set, the density power spectra are much flatter than K41 scaling, and are almost independent of scale ($\propto k^{-0.2}$) for the \texttt{FidMHD} run. The velocity power spectrum is steeper than K41 for hydro, and close to $k^{-1.2}$ scaling for the MHD run. The spectra and structure functions below are averaged from $t=120$ until $200~\mathrm{Myr}$.}
	\label{fig:spectra_fid}
\end{figure}

\begin{figure}
		\centering
	\includegraphics[width=\columnwidth]{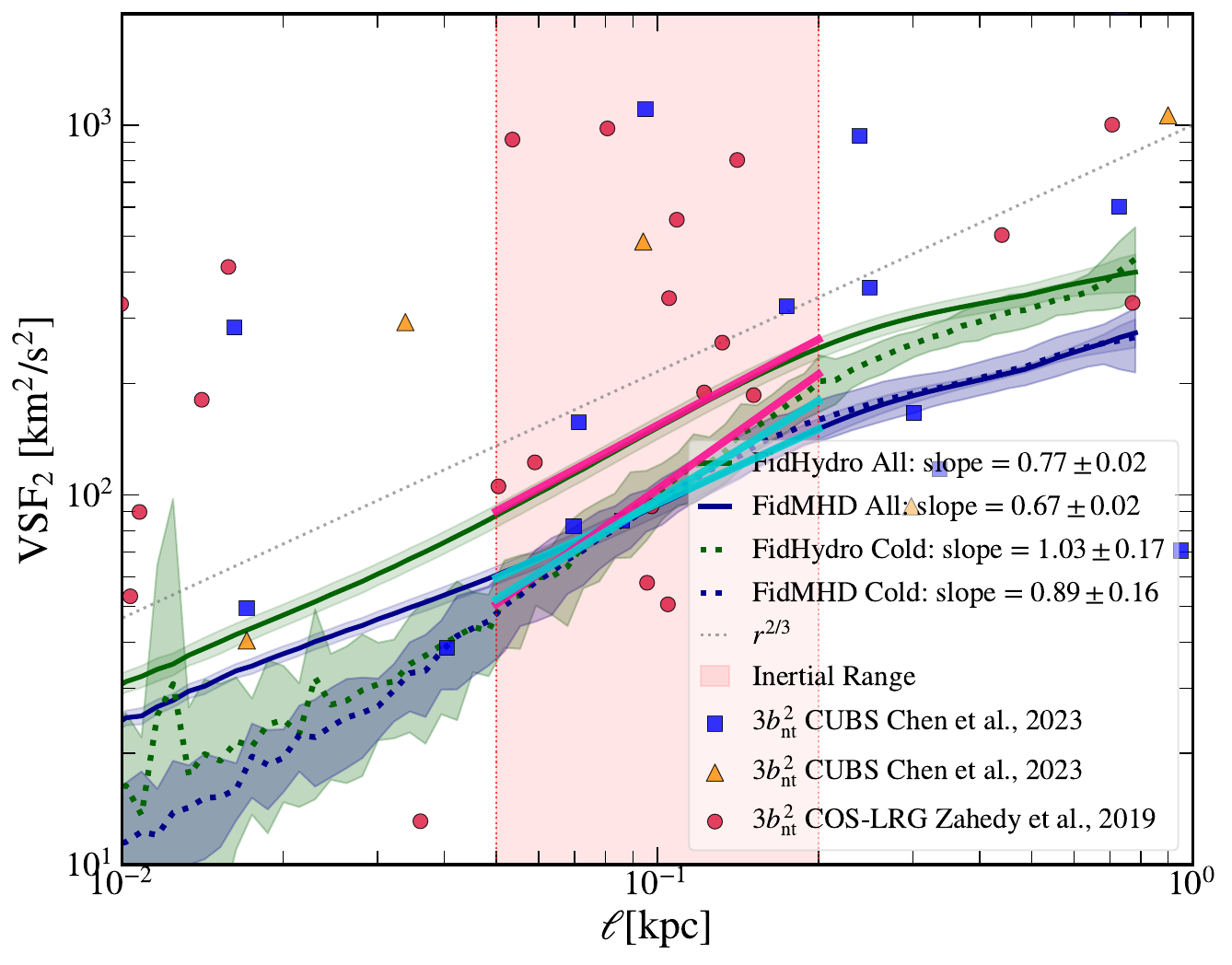}	
	\caption{Velocity structure functions for all gas (solid) and cold gas (dotted) for the \texttt{FidHydro} and \texttt{FidMHD} runs. We also overplot $3b_\mathrm{nt}^2$ inferred from absorption line data from CUBS and COS-LRG samples. Similar to the power spectra results, the $\mathrm{VSF}_2$ is steeper than K41 for Hydro, and close to K41 for the MHD runs. The cold gas $\mathrm{VSF}_2$ has a similar amplitude, but its scaling with separation is slightly steeper compared to the all gas $\mathrm{VSF}_2$ for both runs. }
	\label{fig:vsf_L1}
\end{figure}

We investigate the scale dependence of various statistical properties of our simulations using power spectra and velocity structure functions (VSFs). In \Cref{fig:spectra_LDens}, we present the density and velocity power spectra for the lower average density \texttt{LDens} suite across different snapshots. Subsequently, we show the time-averaged power spectra and VSFs for the  \texttt{Fiducial} suite in steady state in Figures~\ref{fig:spectra_fid} and \ref{fig:vsf_L1}.\footnote{We note here that high-resolution incompressible turbulence simulations using spectral methods have shown that the power spectra slopes only start converging at or beyond $4096^3$ resolution elements \citep{Ishihara2016PhRvF,Yeung2025JFM}. These trends are also seen in ongoing study on compressible large eddy simulations of multiphase turbulence (Grete et al, private communication). Below these resolutions, the power spectrum slope is affected by the bottle-neck effect, which gives rise to a dip in the spectrum at smaller $k$ and a bump at larger $k$. We expect the slopes of power-spectra and structure functions presented in this part of this study to be affected by the dip, and be in general steeper than their converged values.}

For the \texttt{LDens} suite, each snapshot is averaged over a $10~\mathrm{Myr}$ interval centered on the indicated time. The velocity power spectrum in the \texttt{Hydro} run exhibits a slope slightly steeper than the Kolmogorov expectation \citep[$k^{-5/3}$;][]{kolmogorov1941dissipation}, consistent with previous findings for subsonic turbulence \citep{Mohapatra2020}. In contrast, the \texttt{MHD} run shows a significantly flatter velocity power spectrum, with a slope close to $k^{-1.3}$, aligning with results from idealized  MHD simulations (non-radiative, sub-sonic, super-Alfv\'enic turbulence in \citealt{Grete2021ApJ}, multiphase turbulence in \citealt{Fielding2023ApJ} and simulations of ICM turbulence driven by magnetized AGN jet feedback in \citealt{Grete2025ApJ}). \citet{Grete2021ApJ} attribute this flattening to magnetic tension, which mediates large-scale kinetic-to-magnetic energy conversion and suppresses the kinetic energy cascade. 

The density power spectrum in \Cref{fig:spectra_LDens} displays both temporal evolution and sensitivity to phase structure. The green solid lines in the upper panels of \Cref{fig:spectra_LDens} represent the initial seed density perturbations. During the multiphase regime ($15~\mathrm{Myr} \lesssim t \lesssim 75~\mathrm{Myr}$ for \texttt{Hydro}; longer for \texttt{MHD}), the amplitude remains high across scales, but drops sharply as the system transitions to a single-phase state as the dense cold clouds get assimilated into the ambient medium through mixing. In the \texttt{Hydro} run, the slope steepens over time, reflecting faster mixing of smaller clouds due to shorter turbulent mixing times. The \texttt{LDensMHD} run maintains a flat density power spectrum ($\propto k^0$), indicating that multiphase gas introduces density perturbations across all scales. Although the amplitude decreases as cold gas is assimilated into the ambient hot phase, the flatness persists longer in the \texttt{MHD} case due to the suppression of small-scale mixing in the presence of magnetic fields.

In the  \texttt{Fiducial} suite, the steady-state power spectra shown in Fig. \ref{fig:spectra_fid} mirror the behavior observed in the early-time multiphase stages of the \texttt{LDens} runs. The density power spectrum is flat ($\propto k^{-0.4}$ for \texttt{Hydro}; $\propto k^{-0.2}$ for \texttt{MHD}), while the velocity power spectrum is slightly steeper than Kolmogorov for \texttt{Hydro} and close to $k^{-1.2}$ for \texttt{MHD}.

Beyond characterizing the distribution of dense structures and kinetic energy transfer across scales, the density power spectrum has implications for FRB scattering. FRB signals are sensitive to tiny-scale ($\sim\mathrm{AU}$) density fluctuations \citep{Ocker2025ApJ,Mas-Ribas2025arXiv}. Current models, based on single-phase turbulence theory, assume a Kolmogorov-like spectrum to estimate the CGM scattering budget. Our results suggest that multiphase gas can enhance density fluctuations and flatten the spectral slope, potentially increasing the scattering contribution when extrapolated to AU scales.

Finally, we analyze second-order VSFs for all gas and cold gas only, as shown in \Cref{fig:vsf_L1}. VSFs quantify velocity differences as a function of spatial separation and are widely used to study turbulence in the ISM \citep[e.g.,][]{Ha2021ApJ,Ha2022ApJ}, CGM \citep{MChen2024ApJ,MChen2025ApJ}, and ICM \citep{Li2020ApJ,ganguly_nature_2023,Gatuzz2023MNRAS,YLi2023MNRAS,Xrism2025ApJComa}. 

The VSFs for all gas are steeper than the Kolmogorov scaling ($r^{2/3}$), with MHD runs showing flatter profiles than \texttt{Hydro}. On the driving scale, VSFs for all and cold gas are comparable in amplitude, but in the inertial range where the clouds have a shorter mixing time, cold gas VSFs are steeper. Our simulations lack clouds of size $\sim100~\mathrm{pc}$ (see \Cref{fig:vol_rendering_diff_mag_geometry} for a volume-rendering of cold, dense structures in our simulations), so large-scale VSFs reflect inter-cloud velocity differences within a cloud-complex, while smaller scales probe intra-cloud dynamics. This contrasts with our earlier ICM study \citep{Mohapatra2022MNRASa}, where hot and cold phase VSFs differed significantly in both amplitude and scaling. We attribute this to the lower density contrast (by a factor of $10$) between phases in the CGM, which promotes stronger momentum coupling between the phases.

For observational comparison, we overlay non-thermal line broadening ($b_\mathrm{nt}$) versus inferred cloud size ($\ell_\mathrm{cl}$) data from \citet{HWChen2023ApJ}, based on the COS-LRG \citep[e.g.,][]{Zahedy2019MNRAS} and CUBS \citep[e.g.,][]{Zahedy2021MNRAS,Cooper2021MNRAS,ZQu2022MNRASCUBS} samples. Our VSFs for both all gas and cold gas are broadly consistent with the observed non-thermal broadening. However, we have not performed detailed mock spectral analysis to extract intra-cloud velocity spreads. A more thorough comparison is planned for future work; see \citet{Koplitz2023ApJ} for a discussion on line-of-sight turbulence and $b_\mathrm{nt}$.

\section{Local Turbulence Simulations in a Global CGM Context}\label{sec:varying_box_size}

\begin{figure*}
    \centering
    \includegraphics[width=1.8\columnwidth]{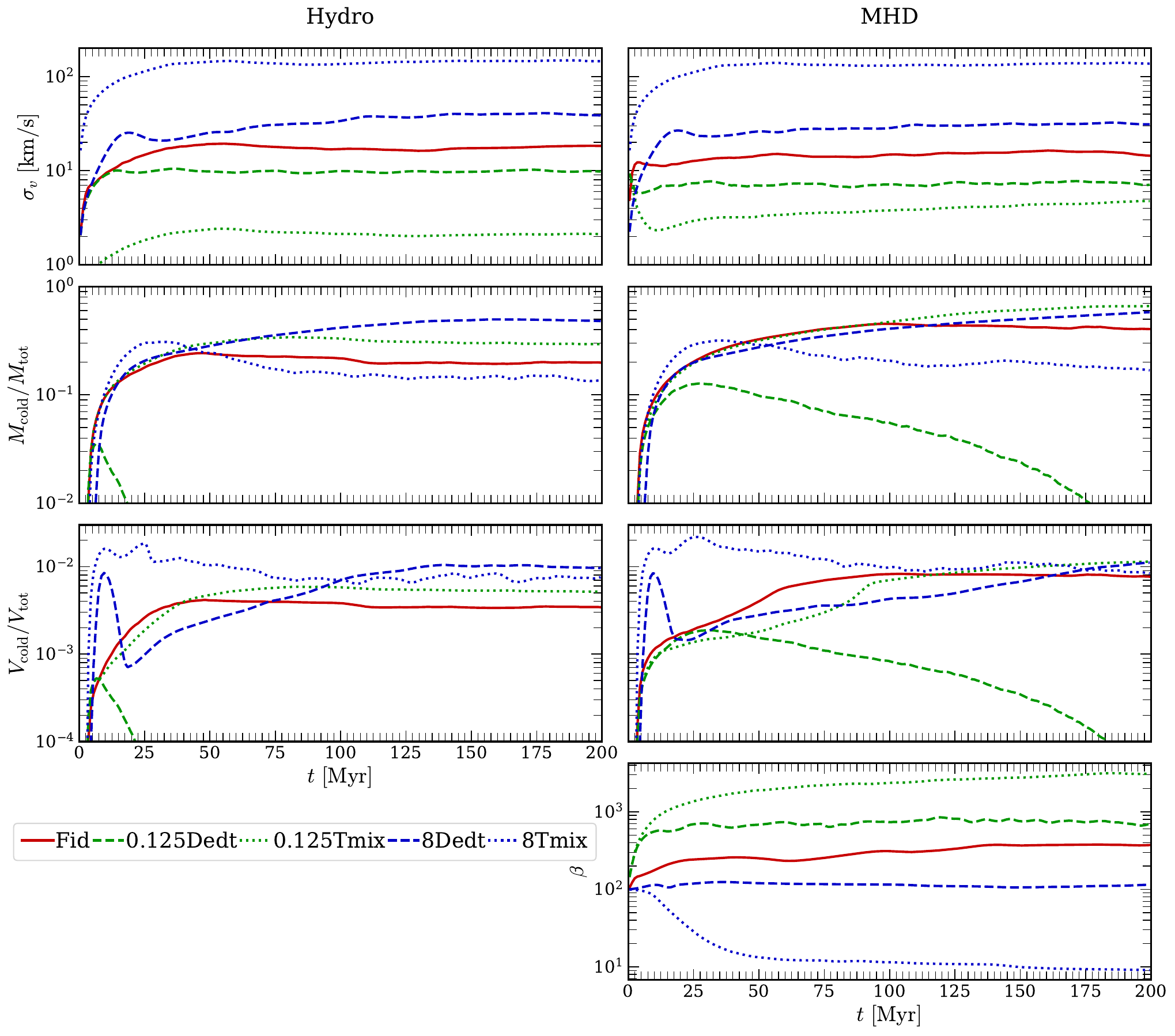}
    \caption{Time evolution of the gas velocity dispersion, and the volume and mass fraction of cold gas for the  \texttt{Fiducial}, fixed \texttt{Dedt}, and fixed \texttt{Tmix} hydro and MHD runs. The \texttt{Dedt} and \texttt{Tmix} runs use box sizes of $8~\mathrm{kpc}$ and $0.125~\mathrm{kpc}$, respectively, compared to the $1~\mathrm{kpc}$ box used in the  \texttt{Fiducial} runs. When the box size is reduced, simulations with matched $t_\mathrm{mix}$ exhibit similar cold gas fractions. In contrast, for fixed \texttt{Dedt} runs, $t_\mathrm{mix}$ decreases with box size, leading to reduced cold gas mass fractions.}
    \label{fig:time_evol_diff_box_size}
\end{figure*}

Idealized turbulence setups, such as those presented in this study, can be challenging to interpret in the broader context of the CGM. To facilitate comparison with galaxy-scale simulations and identify which physical parameters remain consistent across scales, we conduct eight additional simulations: four with a box size of $8~\mathrm{kpc}$ and four with $0.125~\mathrm{kpc}$, including both hydro and MHD variants. All runs use identical initial density perturbation amplitudes, matching those in the  \texttt{Fiducial} suite. The initial conditions for these two runs are described in \S \ref{subsec:vary_box_size_methods}. We briefly outline them again below.

The first set of simulations maintains a constant turbulence energy injection rate per unit volume ($\dot{E}_\mathrm{turb}/\ell_\mathrm{box}^3$) while varying the box size. This approach mimics sub-sampling a CGM region stirred uniformly. Assuming Kolmogorov scaling\footnote{This assumption is a simplification for MHD runs, where the velocity power spectrum is flatter than Kolmogorov, but it serves as a reasonable starting point.}, the total injected energy scales with volume, so increasing (or decreasing) the box size to $8~\mathrm{kpc}$ (or $0.125~\mathrm{kpc}$) increases (or decreases) the net energy input by a factor of $8^3$. With $\dot{E}_\mathrm{turb}/\ell_\mathrm{box}^3$ fixed, the driving velocity scales as $u_{\ell_\mathrm{box}} \propto \ell_\mathrm{box}^{1/3}$. So the ratio $t_\mathrm{cool}/t_\mathrm{mix}$ decreases with decreasing box size. 

The second set of simulations keeps the turbulent mixing time on the box scale ($\ell_\mathrm{box}/u_{\ell_\mathrm{box}}$) approximately constant. This choice is motivated by previous studies suggesting that the ratio $t_\mathrm{cool}/t_\mathrm{mix}$ is a key parameter governing multiphase gas formation, growth, and survival \citep{banerjee2014turbulence,Gronke2022MNRAS,Mohapatra2023MNRAS}. Since the density is unchanged, $t_\mathrm{cool}$ remains constant. The driving velocity is given by $u_{\ell_\mathrm{box}} \propto \ell_\mathrm{box}$.

\Cref{fig:time_evol_diff_box_size} summarizes the results of these simulations compared to the  \texttt{Fiducial} runs (brown solid lines). The left column shows hydro simulations, while the right column shows MHD runs. Simulations with fixed $\dot{E}_\mathrm{turb}$ (labeled \texttt{Dedt}) are plotted with dotted lines, and those with matched $t_\mathrm{mix}^{\ell_\mathrm{box}}$ (labeled \texttt{Tmix}) are shown with dashed lines.

As expected, velocity dispersion $\sigma_v$ varies most significantly in the fixed \texttt{Tmix} runs, and less so in the fixed \texttt{Dedt} cases. However, the volume and mass fractions of cold gas behave differently: runs with matched $t_\mathrm{mix}$ exhibit similar cold gas fractions across box sizes, while those with fixed $\dot{E}_\mathrm{turb}$ show a decline in cold gas content with decreasing box size. This is because $t_\mathrm{mix} \propto \ell_\mathrm{box}^{2/3}$ decreases with box size, increasing $t_\mathrm{cool}/t_\mathrm{mix}$ and the assimilation of cold gas into the hot phase.

These results highlight the importance of $t_\mathrm{cool}/t_\mathrm{mix}$ as a fundamental parameter in multiphase turbulence, potentially more critical than the turbulent heating rate for cold gas survival. 
This observation has important implications for the turbulent, multiphase CGM. In particular, if the whole CGM is uniformly turbulent, it is difficult to sustain cold gas at the observed $\lesssim 10$ kpc scales because of a shorter mixing time at small scales. However, in the presence of locally quiescent and/or dense CGM patches with $t_{\rm cool}/t_{\rm mix} \lesssim 1$, long-lived multiphase gas can exist.

\subsection{Comparison with ICM-like Simulations}\label{subsec:ICM-like simulations}

\begin{figure}
    \centering
    \includegraphics[width=\columnwidth]{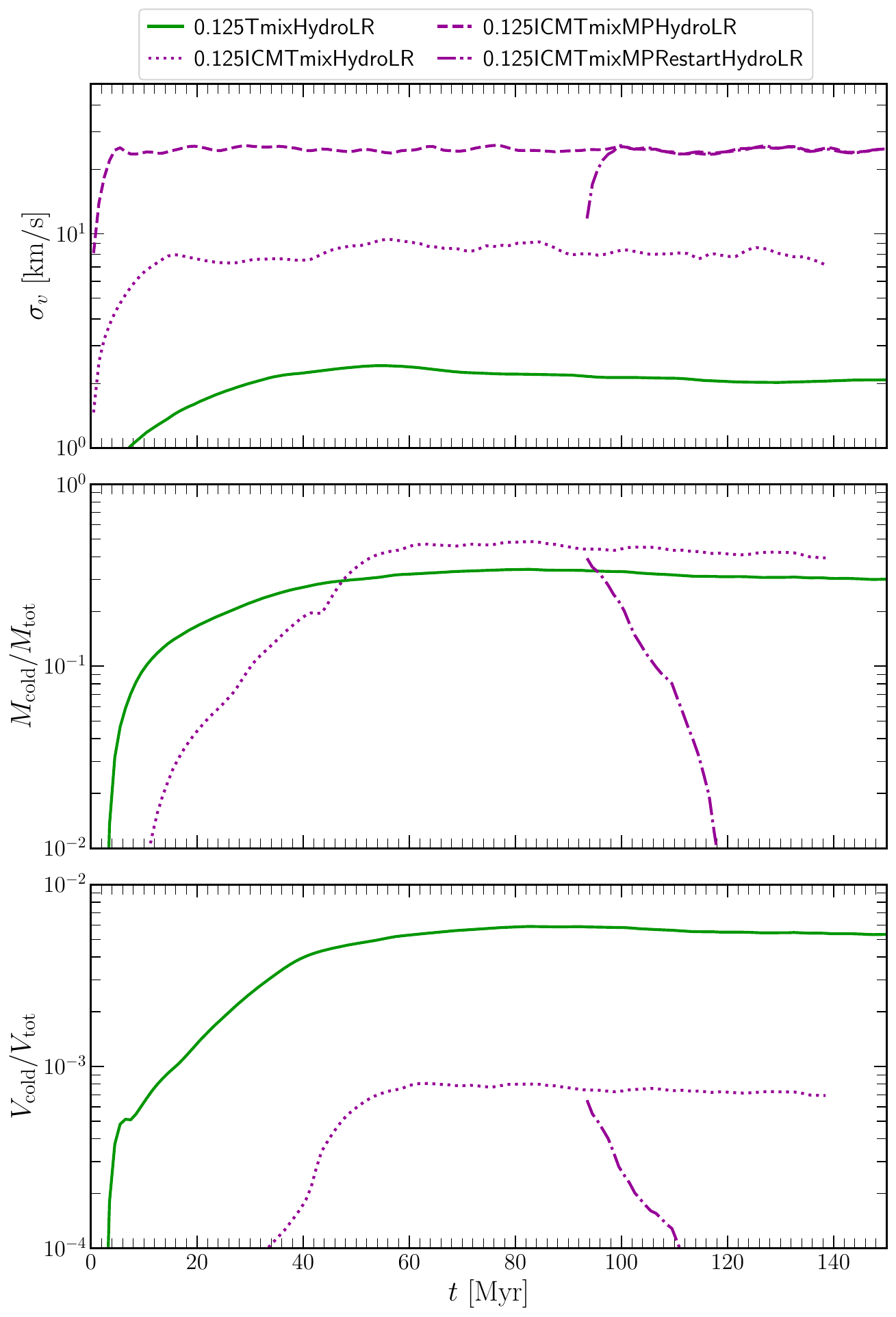}
    \caption{Time evolution of gas velocity dispersion and the volume and mass fractions of cold gas for runs with different density contrasts $\chi$ between hot and cold phases. The \texttt{ICM}-like runs have $\chi$ larger by a factor of 10. When $t_\mathrm{cool}^\mathrm{int}/t_\mathrm{mix}$ is matched between the CGM-like run (\texttt{0.125TmixHydroLR}) and the ICM-like run (\texttt{0.125ICMTmixHydroLR}), the resulting cold gas mass fraction is similar. However, when $t_\mathrm{cool}^\mathrm{int}/t_\mathrm{mix}^\mathrm{multi}$ is matched across runs, stronger driving in the ICM-like cases suppresses cold gas formation or leads to its evaporation. See text for details.}
    \label{fig:time_evol_ICM_lie}
\end{figure}

To assess the impact of density contrast $\chi$ between hot and cold phases, we perform three additional ICM-like runs (Fig.~\ref{fig:time_evol_ICM_lie}; see \S~\ref{subsec:vary_dens_contrast}). These runs adopt a hot-phase density $n=0.03~\mathrm{cm}^{-3}$ and temperature $T=10^7~\mathrm{K}$, with $\chi \sim 1000$, ten times larger than in the CGM-like runs.

We conduct three different runs--for the first run \texttt{0.125ICMTmixHydroLR}, we keep $t_\mathrm{cool}^\mathrm{int}/t_\mathrm{mix}$ fixed between the CGM-like  \texttt{0.125TmixHydroLR} run and the ICM-like \texttt{0.125ICMTmixHydroLR} run, where $t_\mathrm{cool}^\mathrm{int}$ is the cooling time at the geometric mean temperature and density of the two phases. For the second run \texttt{0.125ICMTmixMPHydroLR}, we keep $t_\mathrm{cool}^\mathrm{int}/t_\mathrm{mix}^\mathrm{multi}$ constant across the ICM and CGM-like simulations, where $t_\mathrm{mix}^\mathrm{multi}$ has an extra factor $\sqrt{\chi}$ (see eq.~\ref{eq:t_mix_multi}) and thus it has stronger turbulence with $\sigma_v$ a factor $\sqrt{10}$ times larger than the \texttt{0.125ICMTmixHydroLR} run to account for the difference in the density contrast. For the third run  \texttt{0.125ICMTmixMPHydroLRRestart}, we start it with the initial conditions identical to \texttt{0.125ICMTmixHydroLR}, but once it reaches a steady state, we increase the strength of driving to match that of the \texttt{0.125ICMTmixMPHydroLR} run.

The \texttt{0.125ICMTmixHydroLR} run produces a cold gas mass fraction similar to \texttt{0.125TmixHydroLR}, though with a smaller volume fraction due to the higher $\chi$. The \texttt{0.125ICMTmixMPHydroLR} run forms no cold gas; stronger driving rapidly mixes density perturbations. In \texttt{0.125ICMTmixMPHydroLRRestart}, existing cold gas evaporates once the turbulence driving strength is increased.

At first glance, this suggests $\chi$ does not strongly affect cold gas mass fraction when $t_\mathrm{cool}^\mathrm{int}/t_\mathrm{mix}$ is matched. However, note that $t_\mathrm{mix}$ here is computed on the integral scale, much larger than individual cloud sizes. Clouds in ICM-like runs are smaller (higher mass fraction but lower volume fraction), implying shorter mixing times at cloud scales even if the driving-scale $t_\mathrm{mix}$ is identical. Thus, the commonly used $\sqrt{\chi}$ correction for multiphase mixing may already be captured by smaller cloud sizes. Matching $t_\mathrm{cool}^\mathrm{int}/t_\mathrm{mix}$ at the integral scale therefore yields a similar cold gas mass fraction. In contrast, runs with artificially stronger driving overcompensate for the larger $\chi$, leading to a complete evaporation of the cold phase.

\section{Sensitivity of Multiphase Turbulence 
to Simulation Parameters}\label{sec:sensitiveness_to_sim_params}

In this section, we discuss the effects of the different parameters, such as the magnetic field geometry and compressive forcing on the outcome of our simulations. 
\subsection{Effects of Magnetic Field Geometry}\label{subsec:B-field-geometry}

\begin{figure}
    \centering
    \includegraphics[width=1.\columnwidth]{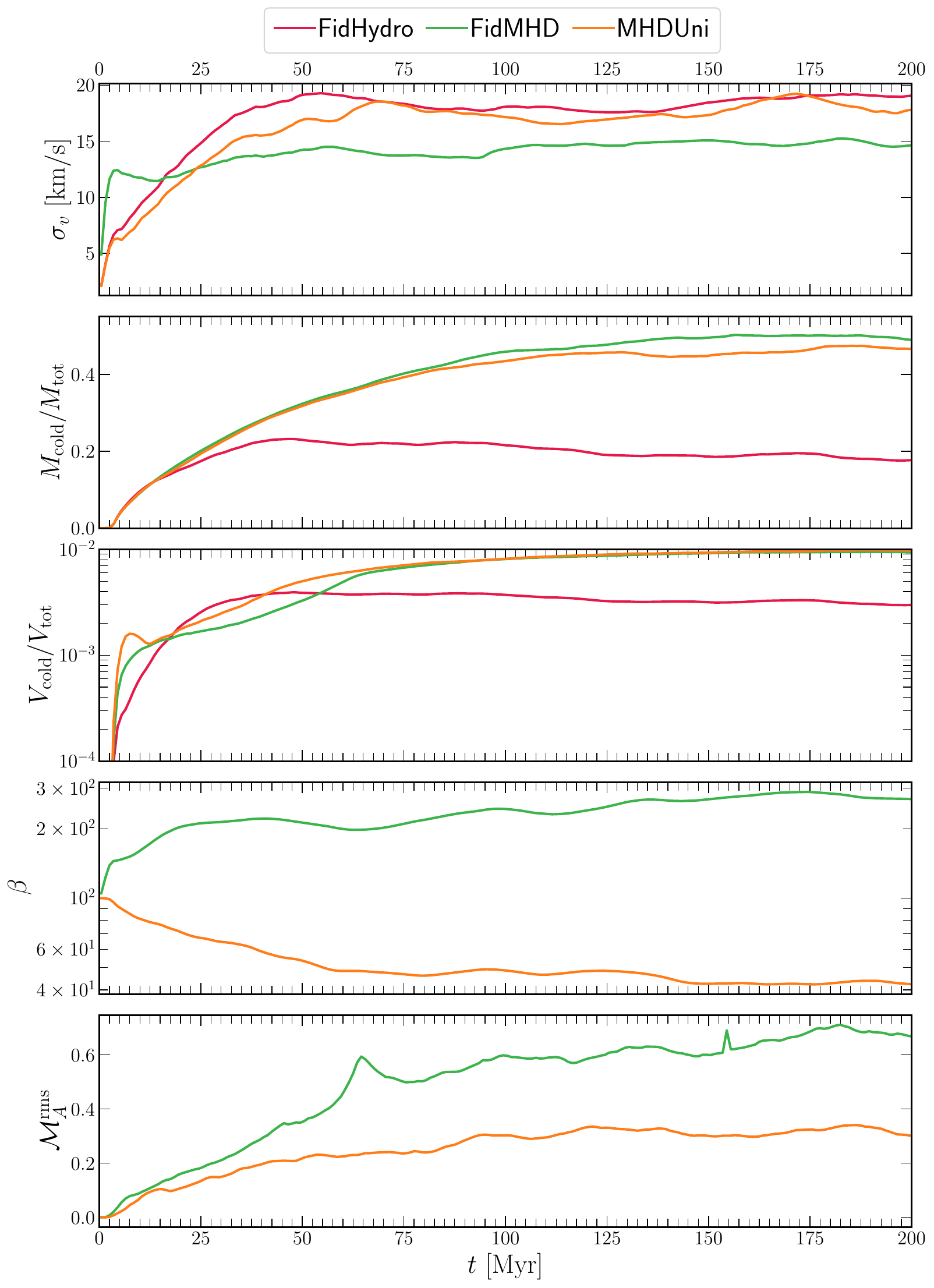}
    \caption{Time evolution of the gas velocity dispersion, and the volume and mass fraction of cold gas for the \texttt{Fiducial} hydro and MHD runs (with random initial magnetic fields), compared to simulations with a uniform magnetic field configuration (\texttt{MHDUni}). While the cold gas fractions are largely insensitive to the initial field geometry, the evolution of plasma beta ($\beta$) and the root-mean-square Alfv\'en Mach number ($\mathcal{M}_A^\mathrm{rms}$) show notable differences.}
    \label{fig:time_evol_diff_mag_geometry}
\end{figure}

\begin{figure*}
    \centering
    \includegraphics[width=2.\columnwidth]{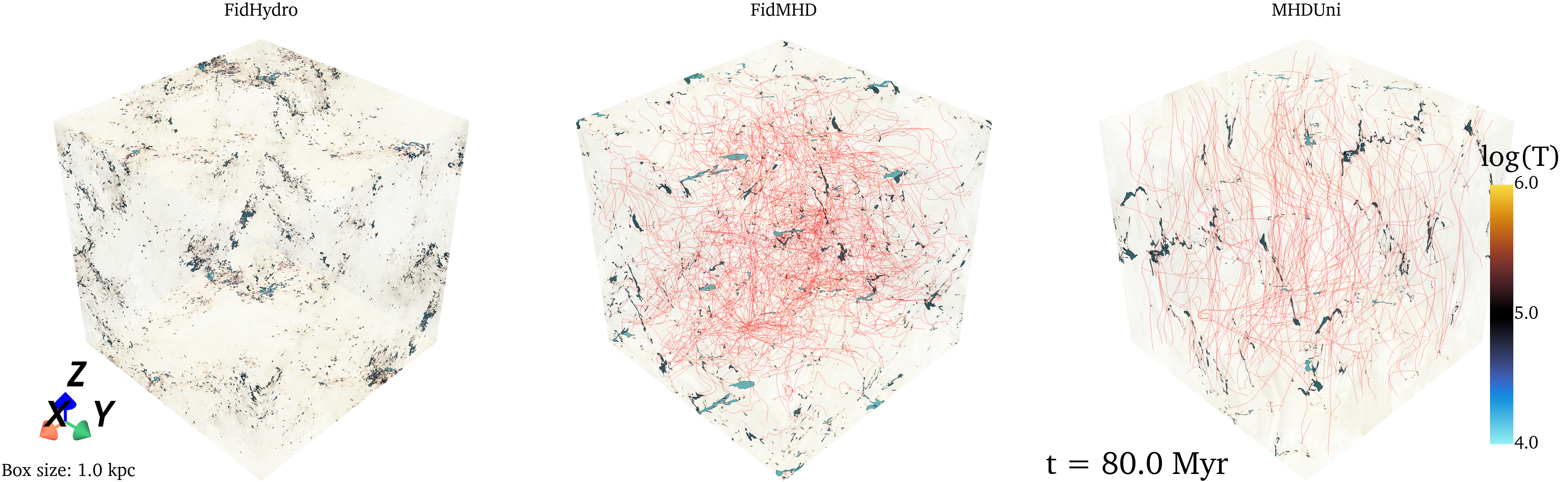}
    \caption{Volume rendering of gas in the \texttt{FidHydro}, \texttt{FidMHD} (with random orientation of initial magnetic fields), and \texttt{MHDUni} (with uniform initial magnetic fields along the $z$-direction) runs at $t=80~\mathrm{Myr}$. The gas opacity is scaled to density, and color encodes $\log_{10}(T)$. Red streamlines in the second and third panels trace magnetic field lines. Cold gas morphology varies significantly across the runs: \texttt{FidHydro} shows fragmentation of cold clouds to small-scales due to shattering, \texttt{FidMHD} exhibits slightly larger clouds due to magnetic pressure support, and \texttt{MHDUni} features larger, coherent cold structures typically aligned with the mean magnetic field, due to suppressed mixing perpendicular to the magnetic fields. An animated version of this figure is available \href{https://youtu.be/I0jYwBYQ7jQ}{here}, as well as in the arXiv submission.}
    \label{fig:vol_rendering_diff_mag_geometry}
\end{figure*}

The strength and configuration of magnetic fields in the CGM remain poorly constrained. Observational evidence from \citet{Bockmann2023A&A} indicates the presence of coherent magnetic fields within the virial radius of galactic halos. Motivated by this, we investigate how the initial magnetic field geometry—specifically a uniform field—affects the evolution of turbulence and cold gas in our simulations.

\begin{figure*}
    \centering
    \includegraphics[width=2\columnwidth]{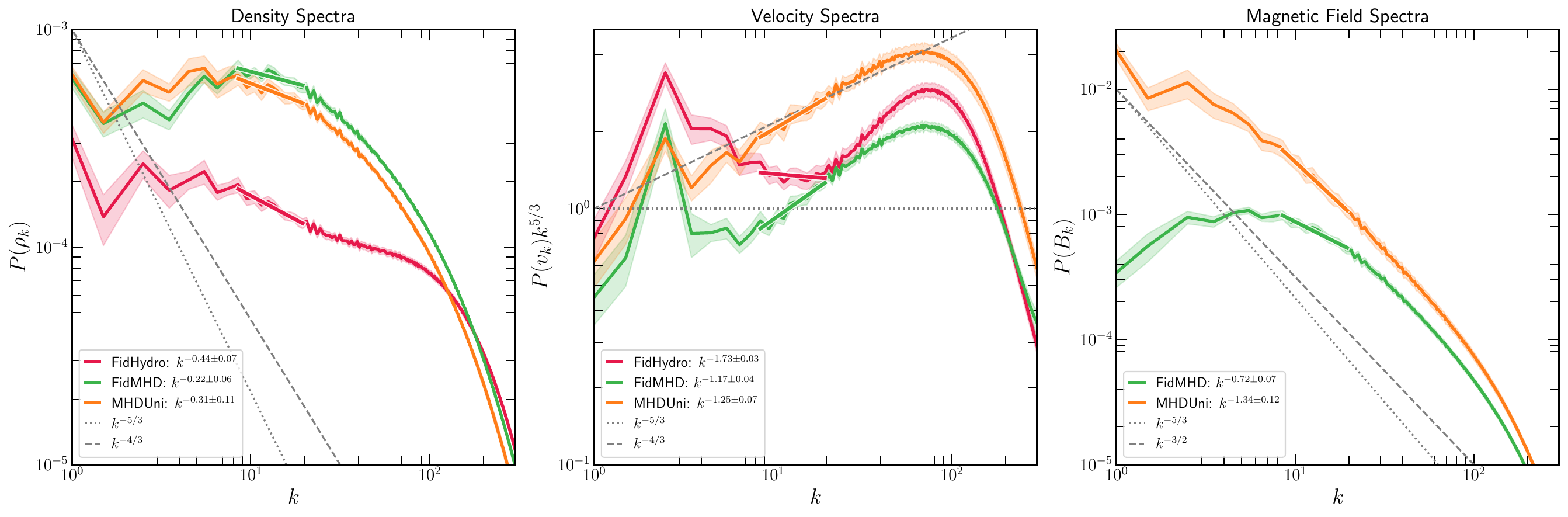}
    \caption{Steady-state power spectra (averaged between $t=120$ to $200~\mathrm{Myr}$) of density, compensated velocity, and magnetic energy for the  \texttt{Fiducial} hydro and MHD runs, and MHD runs with random and uniform magnetic field geometries. The density power spectra are flat on large scales, while the velocity spectra are flatter than Kolmogorov and consistent with $k^{-4/3}$ scaling. The magnetic energy spectra show significant differences on large scales depending on field geometry.}
    \label{fig:spectra_diff_mag_geometry}
\end{figure*}

In \Crefrange{fig:time_evol_diff_mag_geometry}{fig:spectra_diff_mag_geometry}, we compare three setups: \texttt{Fiducial} hydro, \texttt{Fiducial} MHD (with random orientation of initial magnetic fields), and MHD with a uniform initial magnetic field (\texttt{MHDUni}) oriented along the $z$-axis. Details of the initial conditions are provided in \S\ref{subsubsec:mag_fields_init}.

As shown in \Cref{fig:time_evol_diff_mag_geometry}, the velocity dispersion in the uniform field run grows more gradually but reaches a higher steady-state value than in the random field case. Despite this difference, the volume and mass fractions of cold gas are nearly identical between the two MHD runs and remain approximately twice as high as in the hydro simulation. This enhancement is likely due to reduced small-scale turbulent mixing in the presence of magnetic fields.

The initially untangled uniform field undergoes amplification via dynamo processes \citep{Brandenburg2023ARA&A}, leading to a steady increase in magnetic field strength. Consequently, $\beta$ decreases over time, and $\mathcal{M}_A^\mathrm{rms}$ saturates at a lower value compared to the random field case.

Both the presence of magnetic fields and its geometry significantly influence cold gas morphology, as illustrated in \Cref{fig:vol_rendering_diff_mag_geometry}. In the absence of magnetic pressure support, cold clouds in the \texttt{FidHydro} run fragment into smaller structures due to shattering. In contrast, magnetic pressure in the MHD runs suppresses shattering, resulting in larger cold clouds (see also the pressure spread at $T \sim 10^4~\mathrm{K}$ in \Cref{fig:dens_temp_phase_diagram}). Furthermore, in the \texttt{MHDUni} run, cold gas filaments preferentially align with the large-scale magnetic field, reflecting suppressed mixing perpendicular to the magnetic field.

In \Cref{fig:spectra_diff_mag_geometry}, we present the steady-state power spectra of density, velocity, and magnetic energy. The density power spectra for both MHD runs are relatively flat, with slopes of approximately $k^{-0.2}$ for the random field and $k^{-0.3}$ for the uniform field. This flatness reflects the presence of multiphase gas, which introduces density fluctuations across all scales.

The velocity power spectra in both MHD runs are flatter than the Kolmogorov expectation ($k^{-5/3}$), and are closer to the $k^{-4/3}$ scaling reported in recent MHD turbulence studies \citep{Grete2021ApJ,Fielding2023ApJ}. For comparison, \citet{Beattie2024arXiv240516626B} report a steeper slope of $k^{-3/2}$ in high-resolution ($\sim10,000^3$) simulations of supersonic, super-Alfv\'enic, isothermal turbulence, particularly on scales dominated by magnetic fields. Our results show shallower slopes, likely because small scales in our simulations are affected more by the presence of multiphase gas.

The magnetic energy power spectra exhibit the most pronounced differences between the two MHD configurations. While both runs show similar behavior on small scales ($k \gtrsim 20$), the uniform field case displays significantly more power on large scales and a steeper spectral slope. Previous studies report a range of slopes: \citet{Fielding2023ApJ} and \citet{Beattie2023MNRAS} find $k^{-9/5}$, while \citet{Grete2025ApJ} report $k^{-4/3}$. Our results fall within this range and highlight the sensitivity of magnetic energy distribution to initial field geometry. 
Our \texttt{FidMHDUni} simulation with a uniform B can sustain Alfv\'en waves with $k$ as small as $k=1$, but the \texttt{FidMHD} simulation with tangled fields has the largest coherence length of $k=4$, Alfv\'en Waves are sustained at only much smaller scales, leading to a lack of magnetic field fluctuations on the largest scales.

\subsection{Effects of Forcing Prescriptions}
\label{subsec:comp_forcing}
\begin{figure}
    \centering
    \includegraphics[width=1.\columnwidth]{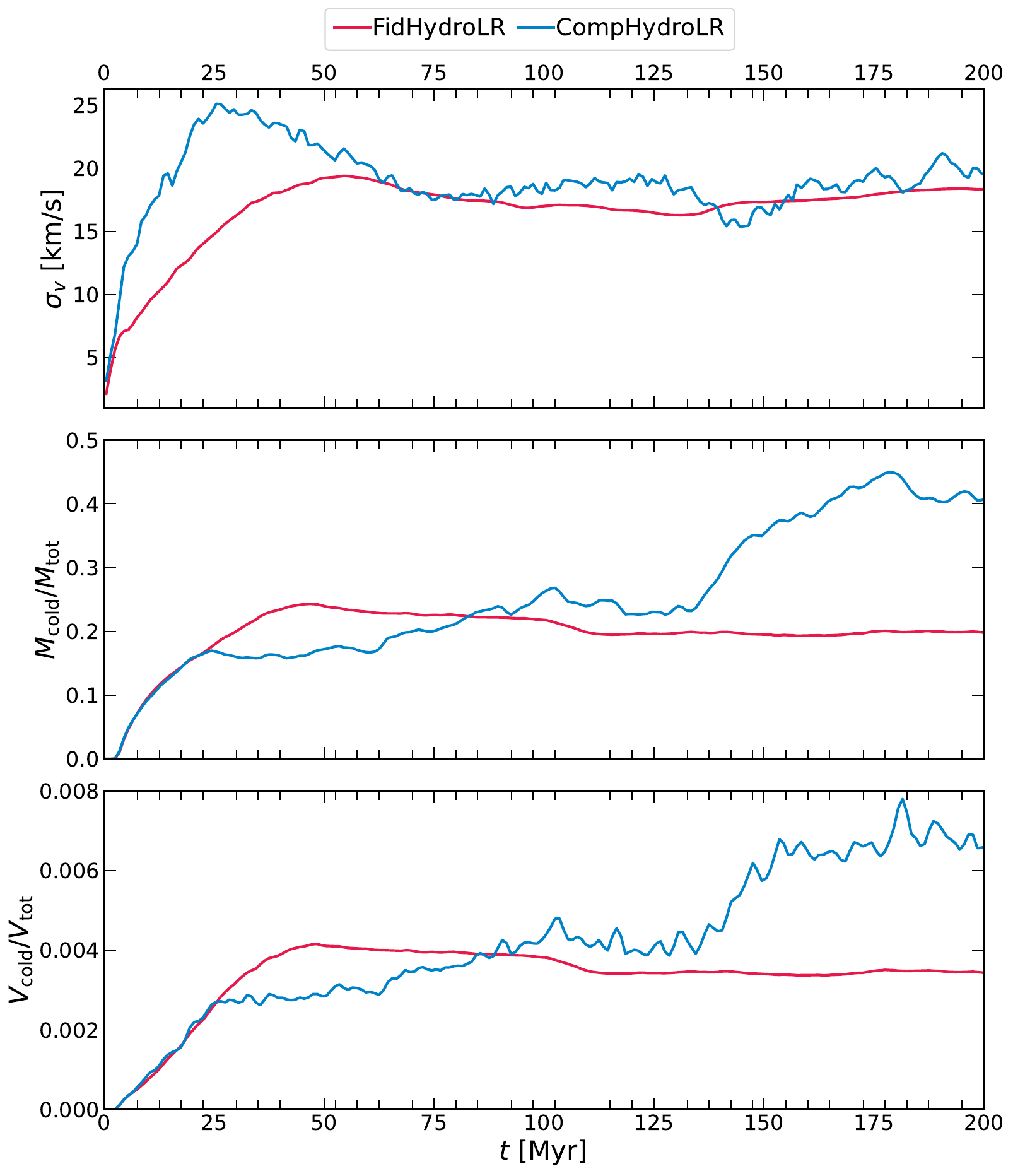}
    \caption{Time evolution of the gas velocity dispersion, and the volume and mass fractions of cold gas for the \texttt{Fiducial} hydro run with solenoidal forcing, compared to the run with compressive forcing (\texttt{CompHydroLR}). Compressive driving produces a larger cold gas fraction than the \texttt{Fiducial} case.}
    \label{fig:time_evol_diff_heating_forcing}
\end{figure}

In this subsection, we examine how the choice of turbulence driving alone affects the formation of cold gas in the CGM. We compare the \texttt{FidHydroLR} simulation, which adopts solenoidal forcing, with the \texttt{CompHydroLR} run, which uses compressive driving. Compressive forcing is relevant for physical situations where turbulence is dominated by shocks or large-scale converging flows, such as those associated with strong gravitational inflows or merger-driven disturbances.

As shown in \Cref{fig:time_evol_diff_heating_forcing}, compressive forcing produces a significantly larger cold gas fraction than solenoidal forcing. This behavior arises because compressive modes generate stronger density contrasts, pushing more gas into thermally unstable regimes and enhancing cooling into the cold phase.

Overall, this comparison highlights that the nature of turbulence driving has a substantial impact on the resulting multiphase structure. Compressive forcing promotes more efficient cold gas formation, underscoring the importance of carefully selecting turbulence driving prescriptions when modeling the CGM.

\section{Caveats and Future Work}\label{sec:Caveats_future_work}

In this section, we outline a few limitations of our current study and discuss directions for future research.

We model the CGM with a fixed metallicity of $0.3Z_\odot$ and assume photoionization equilibrium to compute the cooling function. However, in realistic environments, CGM metallicity is expected to vary with galacto-centric distance, and the ionization state may deviate from both photoionization and collisional ionization equilibrium—particularly in the temperature range $10^4$--$10^{5.5}~\mathrm{K}$. More accurate modeling would require detailed non-equilibrium ionization and species tracking \citep{Buie2018ApJ,Peeples2019ApJ,BuieII2020ApJ}. We also impose a temperature floor of $10^{3.2}\,\mathrm{K}$, below which additional physics such as molecular cooling, shielding, and chemistry become important, and much higher resolution would be required. We plan to address these issues in a future study.

Moreover, we calibrate the turbulent driving velocity in our simulations using the observed non-thermal broadening of absorption lines. However, recent work by \citet{Koplitz2023ApJ} suggests that non-thermal line broadening may not directly correlate with the actual gas velocity dispersion. To address this, we plan to conduct a detailed cloud-by-cloud analysis in future work, generating synthetic absorption spectra to better connect our simulations with observational diagnostics.

In addition, our simulations do not include additional dissipative processes such as thermal conduction, viscosity, and resistivity, and rely on numerical diffusion. While conduction can mediate thermal energy exchange between phases, its effectiveness could be suppressed in cold, magnetized regions \citep{Bruggen2023ApJ,Wang2025arXiv} due to its anisotropic nature. We also do not include the effects of cosmic rays (CRs) in our simulations, which can influence the formation and evolution of cold gas in the CGM \citep{Faucher-Giguere2023ARAA}.  CRs may provide additional pressure support to cold gas and influence its formation and survival \citep{Butsky2020ApJ,Roy2025arXiv}, although recent simulations by \citet{Weber2025A&A} indicate that the impact of CRs depends sensitively on the relative timescales for CR escape and cloud collapse.

We have also ignored stratification of the CGM in our local simulations. Previous work by \cite{sharma2012thermal,Mohapatra2023MNRAS,Wibking2025MNRAS} indicates that stratification plays a role in cold gas formation in a turbulent multiphase medium. The survival of cold clouds could also be affected, and the alignment between the field lines and the stratification could also be important \citep{Kaul2025MNRAS}.

Since we use a periodic box for our simulations, it doesn't allow the gas to adiabatically contract or expand when it overheats or overcools. In most of our simulations, we use an idealized density-dependent heating model that maintains the simulation domain in global thermal balance at every time step, in order to prevent a thermal runaway. We have tested the effect of using a volume-weighted thermal heating model and disabling global thermal balance in \S\ref{sec:appendix_diff_heating_effect}.

Additionally, we fix the driving scale of turbulence to half of the box size in all our simulations, and stir it continuously. In reality, turbulence in the CGM is expected to be driven intermittently across a variety of scales - due to feedback processes originating from the galaxies and motions of infalling substructures. Intermittent and local driving can be possible direction for future studies, and some of these approaches have been explored, for e.g.~\cite{Mohapatra2019} and \cite{Chen2026arXiv} have explored the effects of different driving scales of turbulence, while keeping either $t_{\rm mix}$ or $t_{\rm cool}/t_{\rm mix}$ fixed. \cite{Connor2026ApJ} have explored the effects of localized supernova driving on ISM turbulence. In brief, a smaller driving scale with a similar driving velocity would lead to a smaller $t_{\rm mix}$, and thus make cloud survival less likely. If turbulence/heating is local and not throughout  the CGM, then the cold clouds would be more likely to survive in cooler, undisturbed regions compared to hotter, turbulent regions.  

\section{Summary and Discussion}\label{sec:summary_discussion}

Understanding the multiphase structure of the circumgalactic medium (CGM) is essential for modeling galaxy evolution, yet the distribution and survival of cold gas within the CGM remain poorly constrained. In this study, we have presented a suite of high-resolution magnetohydrodynamic (MHD) simulations to study the formation, structure, and survival of cold gas in the turbulent circumgalactic medium (CGM). Our simulations model a $(1~\mathrm{kpc})^3$ periodic box with initial conditions motivated by quasar absorption-line observations, and resolve the cooling length across all relevant temperatures and densities. We explored both low-density CGM-like environments (\texttt{LDens} runs) and higher-density cloud-complex-like regions ( \texttt{Fiducial} runs), with and without magnetic fields.

\vspace{0.5em}

\noindent\textbf{Key Findings:}
\textbf{Multiphase Gas Formation and Survival:}
\begin{itemize}
    \item All simulations develop a multiphase medium, but cold gas survival depends strongly on the ratio between the cooling and mixing time $t_\mathrm{cool}/t_\mathrm{mix}$. In lower density runs with slower cooling (\texttt{LDensHydro}), cold gas disappears within a few mixing time ($\sim75~\mathrm{Myr}$), while in the presence of magnetic fields \texttt{LDensMHD}, suppressed mixing  extends cold gas lifetime a little further $\sim200~\mathrm{Myr}$  (\Cref{fig:time_evol_fid_ldens}).
    \item In the denser  \texttt{Fiducial} runs, cold gas survives and reaches a steady-state mass fraction of up to $50\%$ in MHD and $20\%$ in hydro, with high area covering fractions ($\sim80\%$), despite occupying only $\sim1\%$ of the volume. Cold and hot phases remain in rough pressure equilibrium even with density contrasts $\gtrsim100$ (\Cref{fig:dens_temp_phase_diagram}).
\end{itemize}

\textbf{Effects of Magnetic Fields:}
\begin{itemize}
    \item Magnetic fields suppress fragmentation and mixing, supporting larger filamentary cold structures compared to their Hydro counterparts (\Cref{fig:projection,fig:slices,fig:vol_rendering_diff_mag_geometry}).
    \item Coherent magnetic fields align with cold filaments and maintain low plasma beta values, reducing thermal interaction between phases and weakening emission from intermediate-ionization species such as \texttt{SiIV} and \texttt{CIV}.
\end{itemize}

\textbf{Scale-dependent Turbulence Statistics and Observational Implications:}
\begin{itemize}
    \item Multiphase turbulence enhances density fluctuations across all scales. In MHD runs, the density power spectrum is flat ($\propto k^{-0.2}$), reflecting the presence of cold gas on a wide range of scales. This has direct implications for FRB scattering, which is sensitive to AU-scale density perturbations (\Cref{fig:spectra_LDens,fig:spectra_fid,fig:spectra_diff_mag_geometry}).
    \item Velocity power spectra are slightly steeper than Kolmogorov in hydro runs and flatter ($\propto k^{-4/3}$) in MHD runs. Velocity structure functions reveal scale-dependent decoupling between cold and hot phases, especially on small scales within cold clouds (\Cref{fig:vsf_L1}). These structure functions can be constrained observationally via emission-line kinematics or inferred from photoionization modeling of absorption features.
\end{itemize}

\textbf{Connecting Local Simulations to Global CGM Models:}
\begin{itemize}
    \item To interpret local turbulence boxes in the context of galaxy-scale simulations, we performed a suite of box-size variation tests ($0.125$, $1$, and $8~\mathrm{kpc}$). As shown in \Cref{fig:time_evol_diff_box_size}, \texttt{Tmix} simulations with matched turbulent mixing times ($t_\mathrm{mix}$) yield consistent cold gas fractions across scales, confirming $t_\mathrm{cool}/t_\mathrm{mix}$ as a robust control parameter for the multiphase CGM.
    \item Simulations with fixed energy injection rates (e.g., \texttt{Dedt} runs) show reduced cold gas content in smaller boxes due to shorter mixing times. 
    This implies that the small-scale ($\lesssim 10$ kpc) cold gas arises in relatively non-turbulent and dense regions in the CGM.
\end{itemize}

Our results demonstrate that cold gas can form and survive under realistic physical conditions in the CGM. However, its survival is sensitive to local thermodynamic and mixing timescales. Future work will need to incorporate additional physics such as anisotropic conduction, non-equilibrium ionization, and generate synthetic absorption spectra through detailed species modeling and cloud-tracking to enable direct comparison with observations.

\begin{acknowledgments} 
RM thanks Eliot Quataert and Prakriti Pal Choudhury for useful discussions.
This work was supported by National Science Foundation (NSF) grants AST-2107872 and AST-2509269, and in part by grant NSF PHY-2309135 to the Kavli Institute for Theoretical Physics (KITP). It work was performed in part at Aspen Center for Physics, which is supported by National Science Foundation grant PHY-2210452.
The analysis presented in this article was performed in part on computational resources managed and supported by Princeton Research Computing, a consortium of groups including the Princeton Institute for Computational Science and Engineering (PICSciE) and the Office of Information Technology's High Performance Computing Center and Visualization Laboratory at Princeton University.
This research used both the DeltaAI advanced computing and data resource, which is supported by the NSF (award OAC 2320345) and the State of Illinois, and the Delta advanced computing and data resource which is supported by the NSF (award OAC 2005572) and the State of Illinois. Delta and DeltaAI are joint efforts of the University of Illinois Urbana-Champaign and its National Center for Supercomputing Applications.
\end{acknowledgments}

\textbf{Software:}\texttt{AthenaK} \citep{Stone2020ApJS,Stone2024arXiv}, \texttt{matplotlib} \citep{Hunter4160265}, \texttt{cmasher} \citep{Ellert2020JOSS}, \texttt{scipy} \citep{Virtanen2020}, \texttt{NumPy} \citep{Harris2020}, \texttt{CuPy} \citep{Okuta2017CuPyA}, \texttt{h5py} \citep{collette_python_hdf5_2014}, and \texttt{astropy} \citep{astropy2018}.


\section{Data Availability}
All relevant data associated with this article is available upon reasonable request to the corresponding author.

\section{Additional Links}
Movies of our simulations are available at this playlist on \href{https://www.youtube.com/playlist?list=PLuaNgQ1v_KMbkypMYmH13_zIhX3_T2J54}{YouTube}.



\bibliographystyle{mn2e}
\bibliography{refs.bib} 



\appendix
\renewcommand\thefigure{\thesection \arabic{figure}} 
\setcounter{figure}{0}   
\setcounter{table}{0}   

\section{Effect of Different Heating Models and Global Thermal Balance}
\label{sec:appendix_diff_heating_effect}
\begin{figure*}
    \centering
    \includegraphics[width=\linewidth]{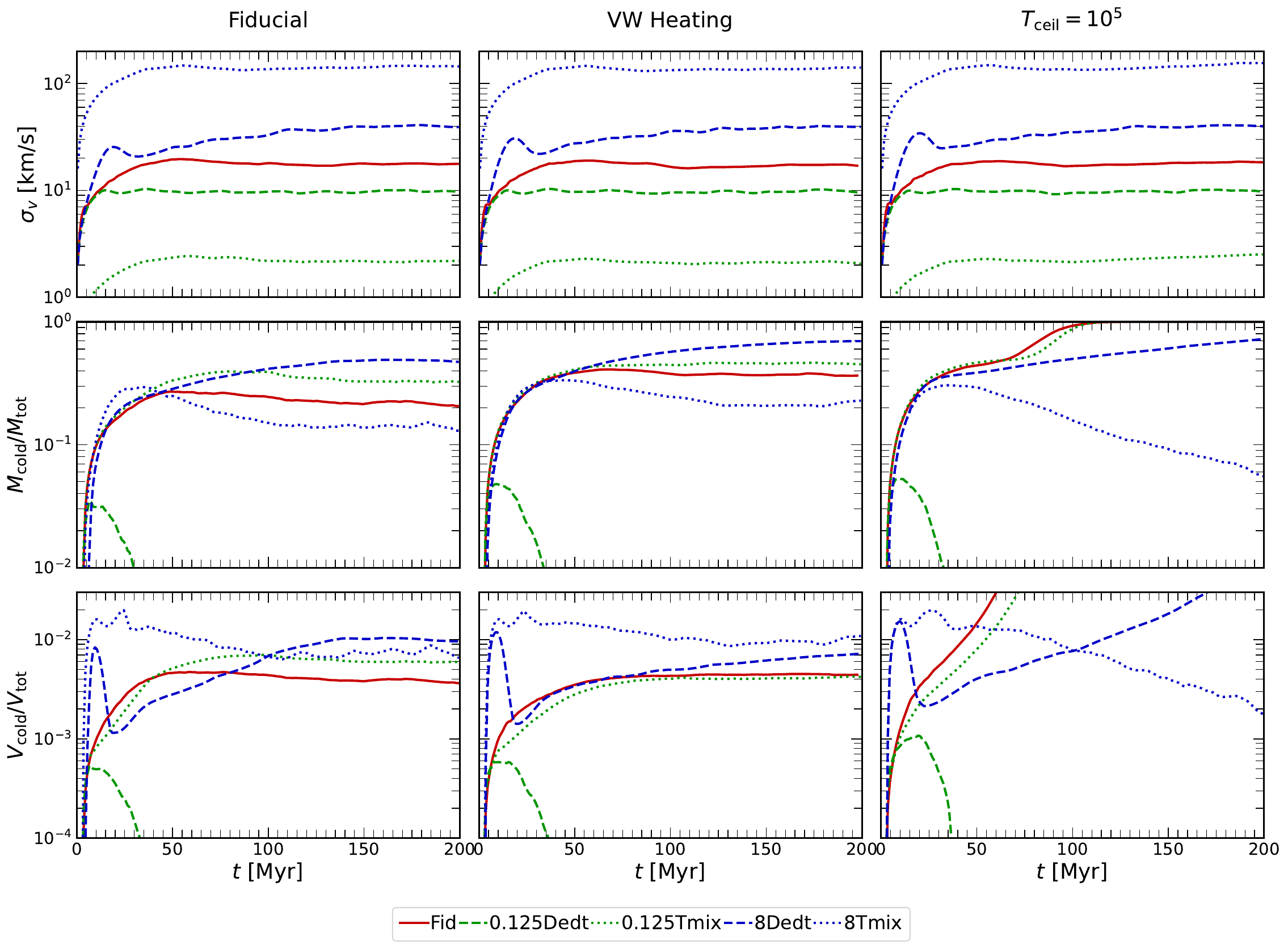}
    \caption{As in \Cref{fig:time_evol_diff_box_size}, we show the time evolution of the velocity dispersion (top row), and the mass and volume fractions of cold gas (middle and bottom rows)  for different runs with a resolution of $256^3$. Each column corresponds to a different heating prescription: density-weighted heating (fiducial; left), volume-weighted heating (middle), and a model in which the system is not maintained in thermal balance, but cooling is disabled for gas above $T_{\rm ceil}=10^5~\mathrm{K}$ (right). Volume-weighted heating yields trends similar to the fiducial model, while the $T_{\rm ceil}=10^5~\mathrm{K}$ model is 
    susceptible to thermal runaway.} 
    \label{fig:appendix_diff_heating_history}
\end{figure*}

Throughout this work we employ a density-dependent idealized heating prescription to prevent global runaway cooling of the simulation domain. In this appendix, we examine how alternative heating and cooling treatments, commonly adopted in the literature to maintain approximately steady thermodynamic states in local periodic simulations, influence our results. We compare these approaches with our fiducial setup in \Cref{fig:appendix_diff_heating_history}. All simulations shown here use a resolution of $256^3$, and we evaluate their impact on the box-size variation experiments described in \S\ref{sec:varying_box_size}, where either $\dot{E}_{\rm turb}/\ell_{\rm box}^3$ (labeled ``Dedt'') or $t_{\rm mix}$ (labeled ``Tmix'') is held fixed while the box size is varied from $0.125$ to $8~\mathrm{kpc}$.

\subsection{Volume-weighted heating}

Our first alternative model is a volume-weighted heating prescription obtained by setting $\alpha_{\rm heat}=0$ (see eq.~\ref{eq:Q_heat}) instead of its fiducial value of $1$. This approach has been used in earlier studies such as \cite{sharma2012thermal,Mohapatra2019} and preferentially heats the low-density, volume-filling hot phase. The corresponding results appear in the second column of \Cref{fig:appendix_diff_heating_history}. Overall, the qualitative behavior closely resembles that of the fiducial model. The fixed-Dedt runs form progressively smaller cold-gas fractions as the box size decreases, with the ``0.125Dedt'' case losing all cold gas by $40~\mathrm{Myr}$. The fixed-Tmix runs yield nearly identical cold-gas mass fractions across all three box sizes.

\subsection{Cooling shutoff above a temperature ceiling}

As a second alternative, we follow previous work \cite[e.g.][]{Gronke2022MNRAS,Das2024MNRAS} and disable radiative cooling for gas with $T>T_{\rm ceil}=10^5~\mathrm{K}$, thereby abandoning global thermal balance. The results for this model appear in the third column of \Cref{fig:appendix_diff_heating_history}. For the first $50~\mathrm{Myr}$, the evolution of the cold-gas mass and volume fractions largely mirrors that seen in the fiducial and volume-weighted cases, including the disappearance of all cold gas in the ``0.125Dedt'' run by $40~\mathrm{Myr}$.  

At later times, however, the system fails to reach a steady state. Although gas hotter than $10^5~\mathrm{K}$ cannot cool directly, mixing with pre-existing cold gas continuously generates intermediate-temperature material in the range $10^4$ to $10^5~\mathrm{K}$, which cools efficiently. As a result, the system undergoes runaway cooling until nearly all gas resides in the cold phase. The exception is the ``8Tmix'' run: after cooling shutoff above $10^5~\mathrm{K}$, turbulent heating exceeds total radiative losses, producing a slow decline in the cold-gas mass fraction at late times.

\section{Effect of Changing Resolution}\label{sec:appendix_resolution_effect}
\setcounter{figure}{0}   
\setcounter{table}{0}   

\begin{figure*}
    \centering
    \includegraphics[width=0.8\linewidth]{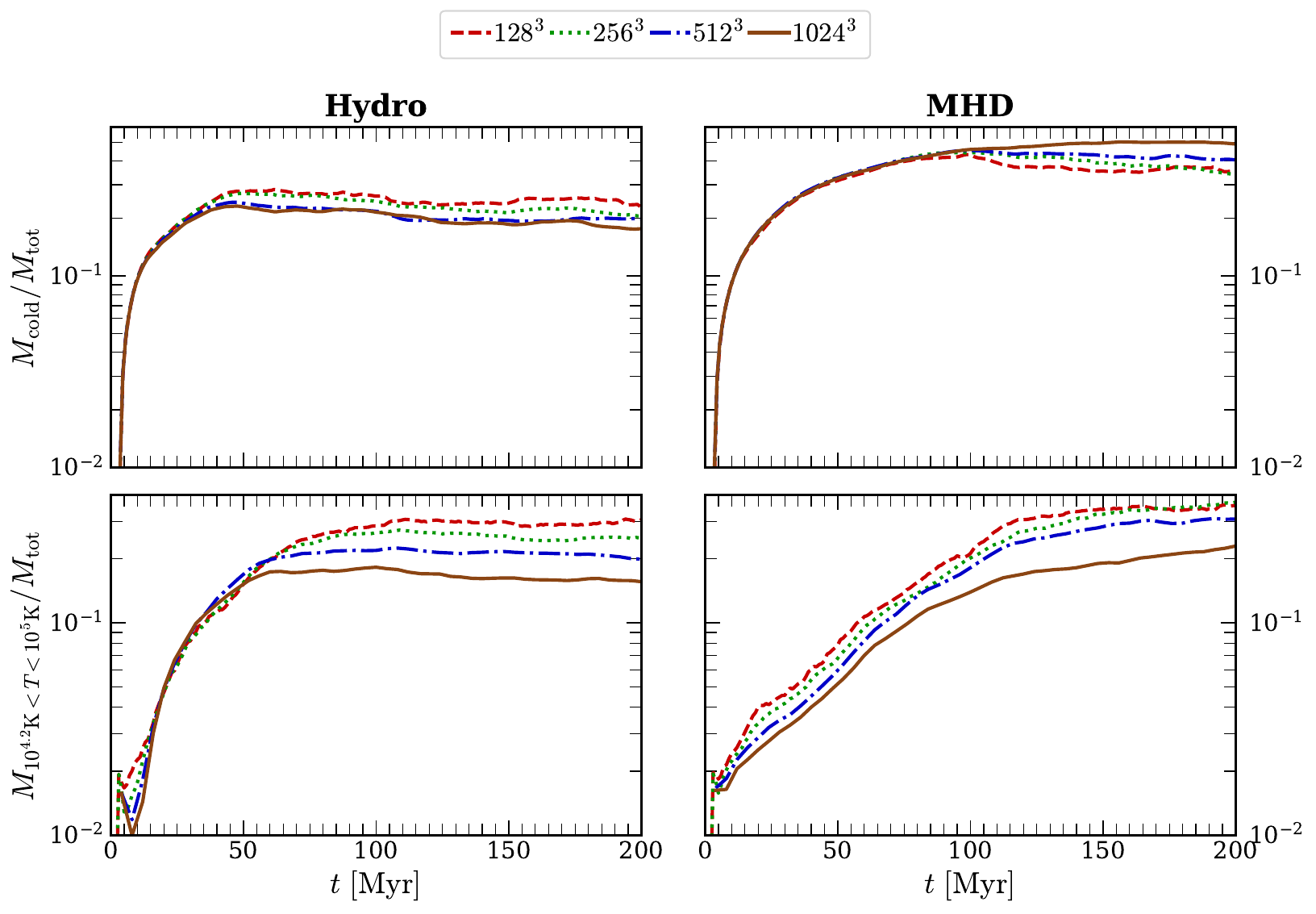}
    \caption{Time evolution of the cold-gas mass fraction for the fiducial hydrodynamic and MHD models at resolutions from $128^3$ to $1024^3$. The cold-gas mass fraction decreases weakly with increasing resolution in the hydrodynamic runs, whereas it increases weakly with resolution in the MHD runs. The mass fraction of intermediate-temperature gas decreases with increasing resolution in both the hydrodynamic and MHD runs.}
    \label{fig:appendix_diff_resolution_history}
\end{figure*}

We now examine the dependence of our fiducial simulations on numerical resolution. Formal convergence would require resolving the relevant viscous and conductive length scales, which remain far below the spatial resolution of the present simulations. We therefore do not claim strict convergence. Instead, we use this appendix to quantify the resolution-dependent trends in the principal diagnostics.

\Cref{fig:appendix_diff_resolution_history} shows the evolution of the cold-gas mass fraction for resolutions from $128^3$ to $1024^3$. The overall differences are modest, but the hydrodynamic and MHD runs show opposite trends. In the hydrodynamic case, the cold-gas mass fraction decreases weakly with increasing resolution. In the MHD case, it increases weakly with increasing resolution. Similar differences between hydrodynamic and MHD resolution trends were reported by \cite{Lancaster2024ApJ} in simulations of wind-blown bubbles, where they were attributed to differences in the excess fractal dimension of the hot-cold gas boundary.

In our simulations, these trends likely reflect a competition between the formation of smaller, denser cold structures and their subsequent mixing. As the resolution is increased, the smallest cold structures become denser (and smaller), as shown in \Cref{fig:appendix_diff_resolution_phase_diagram}. This trend is expected to continue until the cooling length is adequately resolved. In the hydrodynamic runs, these smaller clouds also have shorter cloud-scale mixing times and are therefore more easily mixed into the ambient medium, producing a weak decrease in the net cold-gas mass fraction with resolution. In the MHD runs, small cold structures are more resistant to mixing because magnetic fields suppress small-scale motions. This is consistent with the broader density range of the cold gas in the $512^3$ and $1024^3$ MHD runs in \Cref{fig:appendix_diff_resolution_phase_diagram}. The reduced mixing efficiency in the magnetized case leads to a weak increase in the cold-gas mass fraction with resolution.

The mass fraction of intermediate-temperature gas ($10^{4.2}~\mathrm{K}<T<10^5~\mathrm{K}$), by contrast, decreases with increasing resolution. This suggests that most of this gas is concentrated in thin cooling layers around the cold clouds rather than in an extended volume-filling phase. As the characteristic size of the smallest cold structures decreases with increasing resolution, the total mass in these intermediate-temperature layers also decreases, provided their characteristic densities do not change substantially; see \Cref{fig:appendix_diff_resolution_phase_diagram}. 

\begin{figure}
    \centering
    \includegraphics[width=\linewidth]{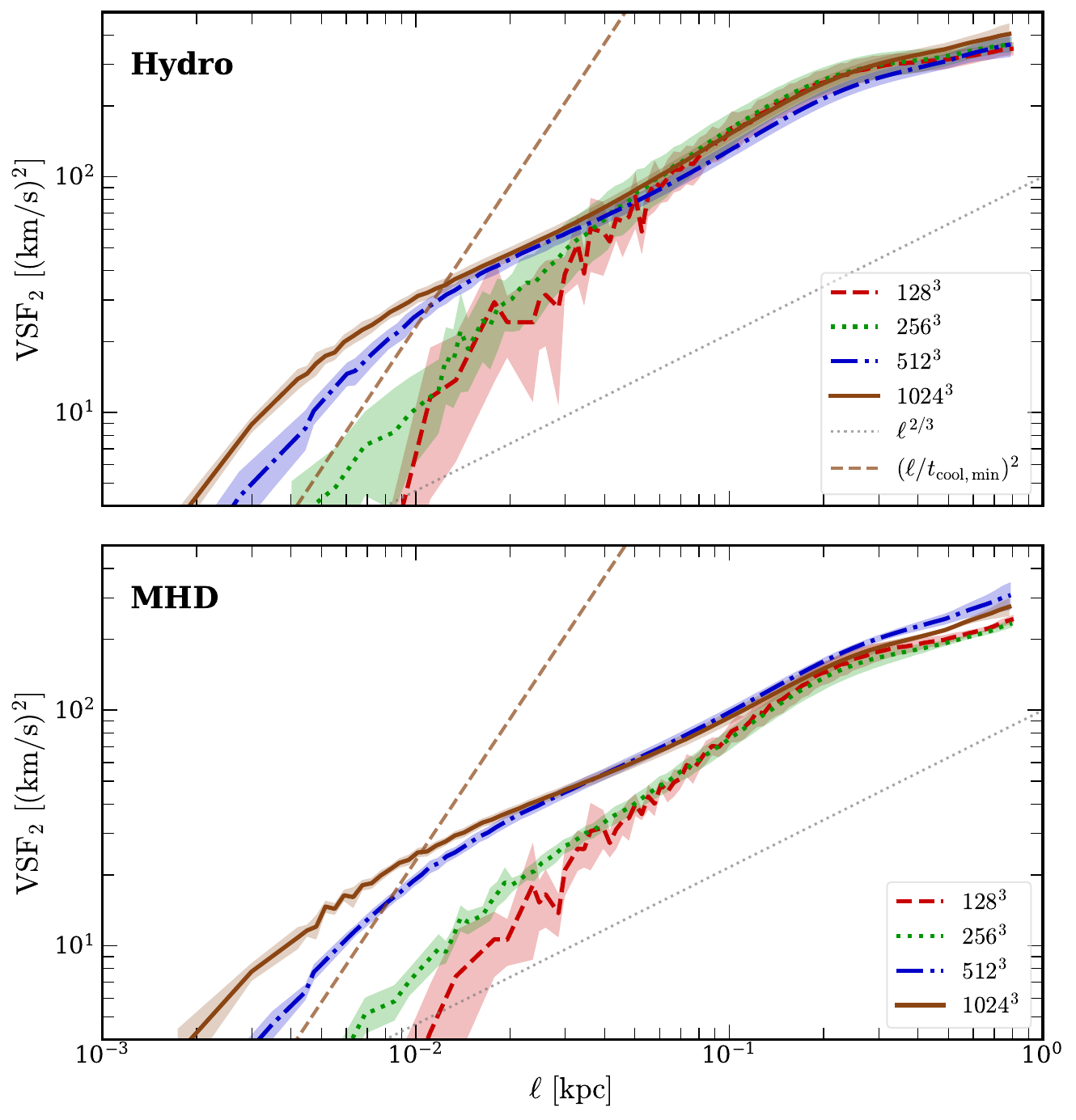}
    \caption{Second-order velocity structure functions for simulations at different resolutions. They continue to evolve with resolution, consistent with the expectation that inertial-range convergence requires substantially higher resolution.}
    \label{fig:appendix_diff_resolution_vsf_all}
\end{figure}

\Cref{fig:appendix_diff_resolution_vsf_all} shows the second-order velocity structure functions at different resolutions. At the resolutions considered here, the measured slopes are still expected to be affected by the bottleneck effect \citep{Ishihara2016PhRvF,Yeung2025JFM}. The approximately flat part of the structure function continues to evolve with resolution.

We also show $(\ell/t_{\rm cool})^2$ in \Cref{fig:appendix_diff_resolution_vsf_all}. According to \cite{Lancaster2026a}, whether mixing is physically resolved on scale $\ell$ depends on whether turbulent diffusion on that scale acts faster than radiative cooling, and the corresponding condition is
\begin{equation}
    {\rm VSF}_2(\ell) > \left(\frac{\ell}{t_{\rm cool}}\right)^2 .
\end{equation}
If this criterion is not met near the grid scale, the intermediate-temperature gas is more likely to arise from numerical averaging across the hot-cold interface than from resolved turbulent mixing. Among our simulations, only the $512^3$ and $1024^3$ runs satisfy this condition for $\ell\lesssim10\Delta x$. Even in these runs, however, the mass fraction of intermediate-temperature gas continues to decrease with increasing resolution, and even higher resolution is required for convergence, as shown in \Cref{fig:appendix_diff_resolution_history}. 

\begin{figure*}
    \centering
    \includegraphics[width=2.0\columnwidth]{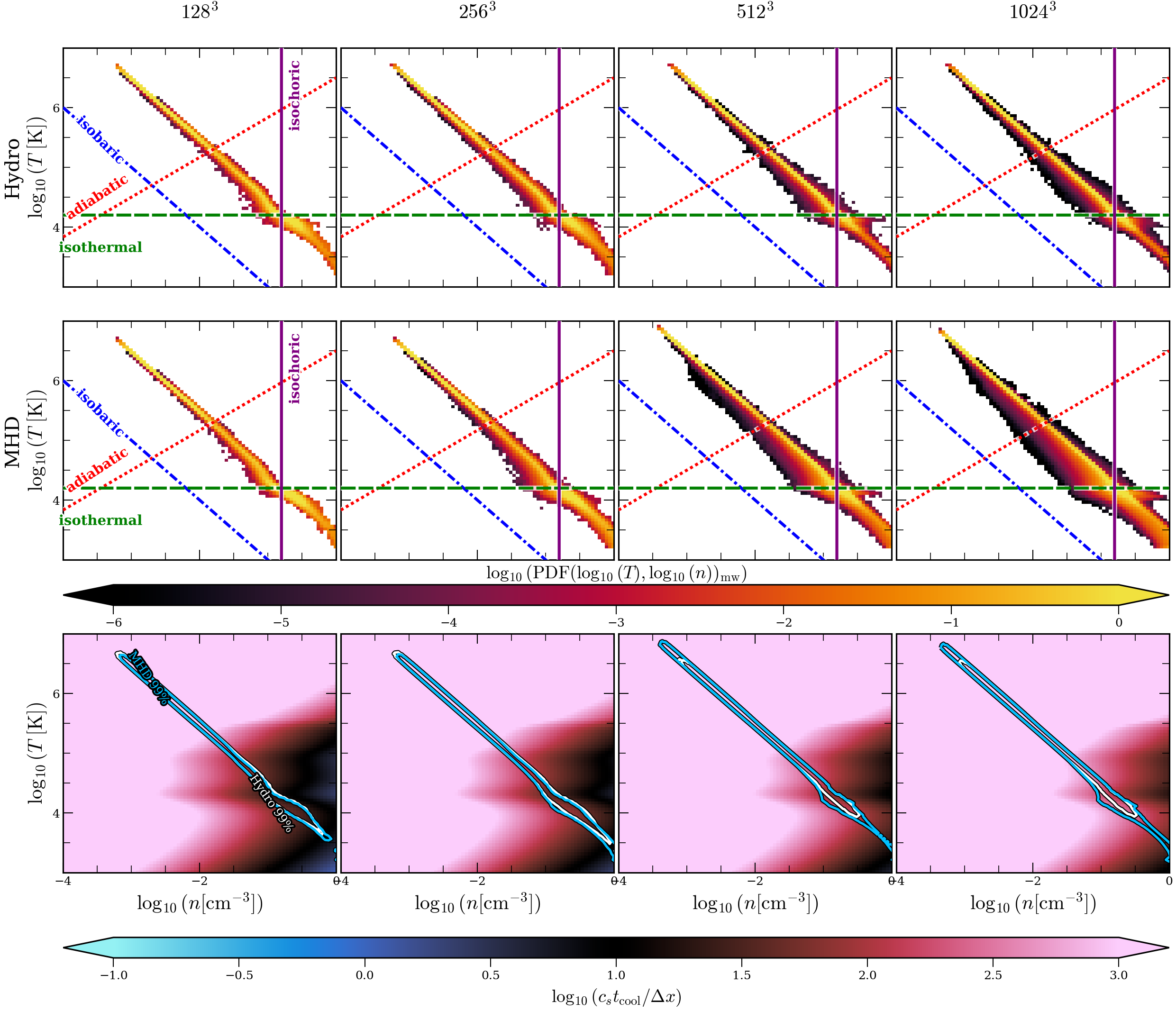}
    \caption{Temperature-density phase diagrams at different resolutions. The top two rows show mass-weighted joint PDFs of temperature and density. The bottom row shows the ratio of the cooling length, $\ell_{\rm cool}=c_s t_{\rm cool}$, to the grid spacing, $\Delta x$. At higher resolution, $\ell_{\rm cool}$ is resolved by at least 10 cells over the relevant phase-space region, and the artificial isochoric feature at $T\lesssim10^5~\mathrm{K}$ becomes weaker. For the MHD run, the spread in thermal pressure at $10^4~\mathrm{K}$ is due to magnetic pressure support in the cold gas.}
    \label{fig:appendix_diff_resolution_phase_diagram}
\end{figure*}

\Cref{fig:appendix_diff_resolution_phase_diagram} shows the temperature-density phase diagrams for the hydrodynamic and MHD runs at different resolutions. In the bottom row, the contours enclose $99\%$ of the gas mass, while the background color shows $c_s t_{\rm cool}/\Delta x$. At low resolution, gas with $T<10^5~\mathrm{K}$ has unresolved cooling length and appears artificially under-pressured relative to the hot phase. At higher resolution, where $\ell_{\rm cool}/\Delta x \gtrsim 10$ over the relevant region of phase space, this feature is reduced and the cold and hot phases are closer to pressure balance.

The maximum and mean densities of the $10^4~\mathrm{K}$ gas increase with increasing resolution. Similar behavior has been reported in simulations of turbulent radiative mixing layers \citep{Fielding2020ApJ} and cloud-wind interactions \citep{Abruzzo2024ApJ}. When the cooling length is under-resolved, the phase diagram shows a spurious, nearly vertical isochoric branch around $n\simeq0.1~\mathrm{cm}^{-3}$. This feature becomes progressively weaker as the cooling length is better resolved. Although a pressure dip at the hot-cold interface can also physically arise from substantial compressive stress or from a strong bulk flow of hot gas into the cold phase \citep{Sharma2025,Lancaster2026a}, such an effect should not depend on resolution, and we therefore do not believe that it is responsible here. In the MHD runs, however, a broad pressure distribution at $T \sim 10^4~\mathrm{K}$ persists even at high resolution. This spread is not a numerical consequence of unresolved cooling, but instead reflects magnetic pressure support within the cold gas.

\end{document}